\documentclass[]{pracamgr}

\usepackage{feynmp-auto}
\usepackage[T1]{fontenc}
\usepackage[utf8]{inputenc}	
\usepackage{amsmath}
\usepackage{array}
\usepackage{caption}
\usepackage{latexsym}
\usepackage{booktabs}
\usepackage{mathrsfs}
\usepackage{graphicx}
\usepackage[usenames,dvipsnames,svgnames,table]{xcolor}
\usepackage{float}
\usepackage{multicol}
\usepackage[a4paper]{geometry}
\usepackage{setspace} 
\usepackage{slashed}

\newcommand{\te}{\renewcommand{\arraystretch}{1.4}}
\newcommand{\mc}[3]{\multicolumn{#1}{#2}{#3}}
\numberwithin{equation}{section}

\author{Mateusz Duch}
\title{Effective Operators\\ for Dark Matter Interactions}

\tytulpol{Efektywne operatory dla oddziaływań ciemnej materii}

\kierunek{PHYSICS}

\opiekun{prof. dr hab. Bohdan Grz\k{a}dkowski\\
  Institute of Theoretical Physics,\\
  Faculty of Physics, University of Warsaw.
  }

\date{July 2014}

\nralbumu{290727}
\dziedzina{13.2 Fizyka}

\klasyfikacja{Fizyka}

\begin{document}
\maketitle
\begin{abstract}
The aim of this thesis is to determine possible interactions between the Standard Model with right-chiral neutrinos and a sector of dark matter 
composed of a real scalar, left- and right-chiral fermion and a vector particle, which are singlets of the Standard Model gauge group. 
In the introductory chapters we review the evidence for dark matter and methods of its detection. In the main part, we present the complete 
list of gauge singlet operators of dimension not greater than 4, that consist of the standard sector fields. We obtain also an analogous 
list for the dark matter particles. Using both of them, we find the possible interactions between the two sectors, firstly for the 
renormalizable theories, then within the effective theory framework with operators up to dimension 6, imposing the stability symmetry 
condition on each dark matter field. We consider the problem of generation for the effective operators in underlying, renormalizable gauge 
theory that includes the Standard Model. It is shown, which of the operators can be generated in tree-level approximation and which require a loop.

%
\end{abstract}

\tableofcontents
\chapter*{Introduction}
\addcontentsline{toc}{chapter}{Introduction}

Understanding the nature of dark matter (DM) is one of the major problems of contemporary cosmology, astrophysics and particle physics. 
Since 1930s,  when the problem of ``missing matter'' first arose in the context of galaxy rotation curves, 
despite major advances in studies of the~universe and the role of dark matter in its formation, evolution and present behaviour,
we are still unable to find out how dark matter incorporates into the particle physics framework.

Numerous theories beyond the Standard Model (SM) with sensible dark matter candidates have been proposed. Among them are:
supersymmetric particles, additional scalars in the~extensions of the SM Higgs sector, axions, dark matter in technicolour or extra dimensions theories, 
to mention only the most intensely studied \cite{bertone}. Nonetheless no traces of new physics were found, neither at the LHC, nor at other particle colliders,
while scalar particle, which fits the Standard Model predictions for Higgs boson was discovered. 
Furthermore experimental efforts leading to direct detection of dark matter particles are still inconclusive \cite{bergstrom}.

As a fundamental theory of dark matter is not known, it is useful to approach the problem in~the model-independent manner,
in the context of an effective field theory. Such theory describes only low-energy processes disregarding possible UV completions. Integrating out heavy
modes in the fundamental theory, one obtains an effective Lagrangian which contains infinite number of operators suppressed by increasing powers
of $\Lambda$ - the scale of high-energy modes. This work aims to present the classification of effective operators 
that parametrize possible interactions between the Standard Model and general dark matter candidates: scalar, fermion or vector particles.

In the first two chapters we present a short review of increasing evidence for dark matter  and methods of its detection. Then in the main part
of the thesis we firstly present the Standard Model with right-chiral neutrinos ($SM\nu_R$) and find the list of all gauge singlets operators
up to dimensions 4 in different Lorentz representations (tab. \ref{tablesm}). Then we complete a similar list (tab. \ref{tabledm}) 
for particles in the dark sector, that are assumed to be singlets of the SM gauge group, but without enforcing any stabilising symmetry. 
In the next part we use the previous results to obtain a list of possible interactions between $SM\nu_R$ and dark sector, firstly 
for the renormalizable theories with operators up to dimension 4, then within effective field theory up to dimension~6 with
the assumption of the stabilising $Z_2$ symmetry for every dark matter particle~(tab. \ref{tablesmdm}).
We focus on the problem of mass generation for vector fields and indicate differences and similarities between the Stuckelberg and the Higgs mechanism. Finally
we handle the issue of the effective operators generation within a general renormalizable gauge theory and find out, 
following classification of suppression for effective operators in the SM \cite{artz}, which operators are generated by tree and which by loop graphs.

\chapter{The evidence for dark matter}
The basic assumption of the contemporary cosmology is that the homogeneous, isotropic and flat Universe is governed by the laws of 
General Relativity. Numerous experiments give evidence that the Universe is filled not only with the known baryonic matter, but also with  two other constituents:
dark matter and dark energy. Dark energy is the form of energy  with negative pressure, that homogeneously permeates the space causing
the accelerating expansion of the Universe. Its exact nature and origin are still unclear, however
it can be described by Einstein's equations with cosmological constant. On the other hand, observations indicate that dark matter behaves 
like the ordinary matter. It interacts gravitationally and has tendency to cluster, 
but as its name suggests, it does not emit or scatter light and therefore is difficult to detect by standard astronomical methods.

The commonly adopted model of cosmology is the $\Lambda\text{CDM}$ model, which includes dark energy in the form of cosmological constant and
cold dark matter, that moves slowly compared to~the speed of light. The $\Lambda\text{CDM}$ is in a good agreement with contemporary,
precise observations. They indicate that only $5\%$ of the Universe energy is in the form of known baryonic matter, the rest 
is described by the dark components. Currently the best experimental value for dark energy $\Omega_\Lambda$, 
dark matter $\Omega_c$, and baryon $\Omega_b$ energy density 
(calculated with respect to $\rho_c$ - the current critical density, $\Omega=\rho/\rho_c$) comes from data produced by Planck and  WMAP satellite survey 
of the cosmic microwave background (CMB)~\cite{planck}
\begin{equation}
 \Omega_\Lambda=0.683\;\;\;\;\;\;\;\;\;\Omega_c=0.268,\;\;\;\;\;\;\;\;\;\Omega_b=0.049.
\label{values}
\end{equation}

The evidence for dark matter has various character and spans from the galaxy scale to the Universe as a whole. From various observations
and theoretical models we can infer the essential properties of
dark matter, namely it is electrically neutral (therefore non-luminous), massive, non-baryonic, non-relativistic (cold) and
weakly interacting with the ordinary matter. Moreover dark matter must be stable or decay with a lifetime much larger than the age of~the Universe.
This requirements cannot be fulfilled by any known particle. Many reasonable candidates
beyond the Standard Model were proposed. To became one of them, a particle must pass a ten-point test \cite{test}.

The problem of "missing matter" was first noted by Franck Zwicky in 1933. He was observing the Coma cluster and using viral theorem found that
it contains insufficient amount of luminous matter to explain distribution of galaxy velocities. Similar conclusions can be reached 
by examination of the rotational curves for single galaxies. The measurements exhibit (fig. \ref{rotationcurve}) that the galaxy rotation curves 
on large distance remain flat, while the Newton's gravitational law provides the following expression 
for circular velocity $v(r)$
\begin{equation}
 v(r)= \sqrt{\frac{GM(r)}{r}},
\end{equation}
where the mass of matter in the radius $r$ is given by
\begin{equation}
  M(r)=\int_0^{r} 4 \pi \rho(r) r^2 dr.
\end{equation}
This means that for large $r$, where the density of luminous matter is negligible and $M(r)$ is nearly constant, 
the rotation curves should decrease as $r^{-1/2}$, but the observed shape advocates the existence of dark matter galaxy halos with 
$\rho(r)\propto r^{-2}$, that gives $M(r)\propto r$.



It turns out that the observed rotational curves could be described by the modification of Newtonian gravity \cite{bertone}, nevertheless the observation
of cluster collisions are hardly explicable without the existence of dark matter. The clearest evidence is given by the famous Bullet Cluster 
(fig. \ref{bulletcluster}) that consists of matter perceivable in the visible and X-ray spectra
and weakly interacting dark halos detected by gravitational lensing, that were not decelerated during collision and separated from baryons.

\noindent\begin{minipage}[t]{0.48\textwidth}
\centering
\includegraphics[scale=0.6]{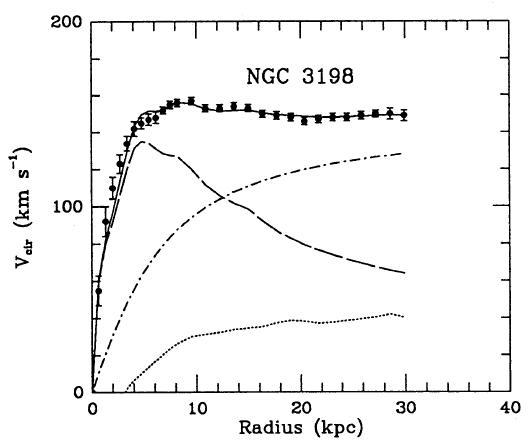}
\captionof{figure}{\small\label{rotationcurve}Rotation curve of NGC 3198 galaxy. Solid line is fitted to experimental values.
Dashed curves refers to the presence of only a visible component, 
the dotted is for a gas, and the dash-dot for a dark halo \cite{ngc}.}
\end{minipage}
\hspace{0.5cm}
\begin{minipage}[t]{0.48\textwidth}
\centering
\includegraphics[scale=1.2]{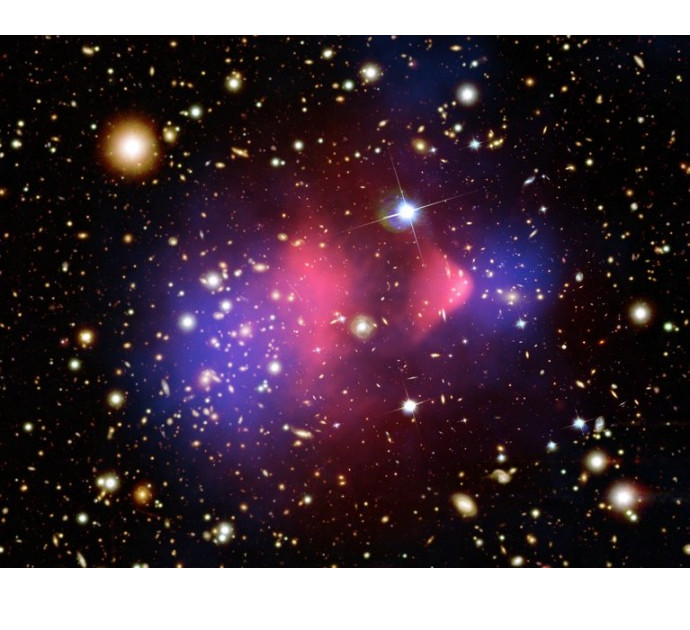}
\captionof{figure}{\small\label{bulletcluster}The Bullet Cluster with the luminous matter (red) and two dark matter halos (blue). 
Composite credit for X-ray, optical, and lensing map: NASA/CXC/CfA/STScI/D.Clowe et al. ESO/Magellan/U.Arizona/  M.Markevitch et al.} 


\end{minipage}

\vspace{1cm}
We can also infer dark matter properties from the observations of the universe on larger scales. In the scale structure 
formation not only the density but also type of dark matter particle is important \cite{bergstrom2000}. In the case of the light particles, moving
with relativistic velocities, dark matter would escape from the overdense galaxy-size regions preventing formation of the smaller structures.
This results in a top-down hierarchy of structures with the smaller formed by the fragmentation of the larger. Such behaviour is in the contradiction with the 
galaxy distributions observed  at very high redshift. It favours the existence of cold dark matter with a mass of order GeV or higher. 
On the other hand, the particles moving 
with non-relativistic velocities can concentrate on smaller scales. It leads to the opposite structure formation with the smaller
 ones merging into the larger.
This viewpoint is also supported by the N-body simulations like the Millennium project \cite{millenium} or recently Illustris \cite{illustris}.

\begin{figure}[htb!]
\begin{center}
\includegraphics[scale=0.4]{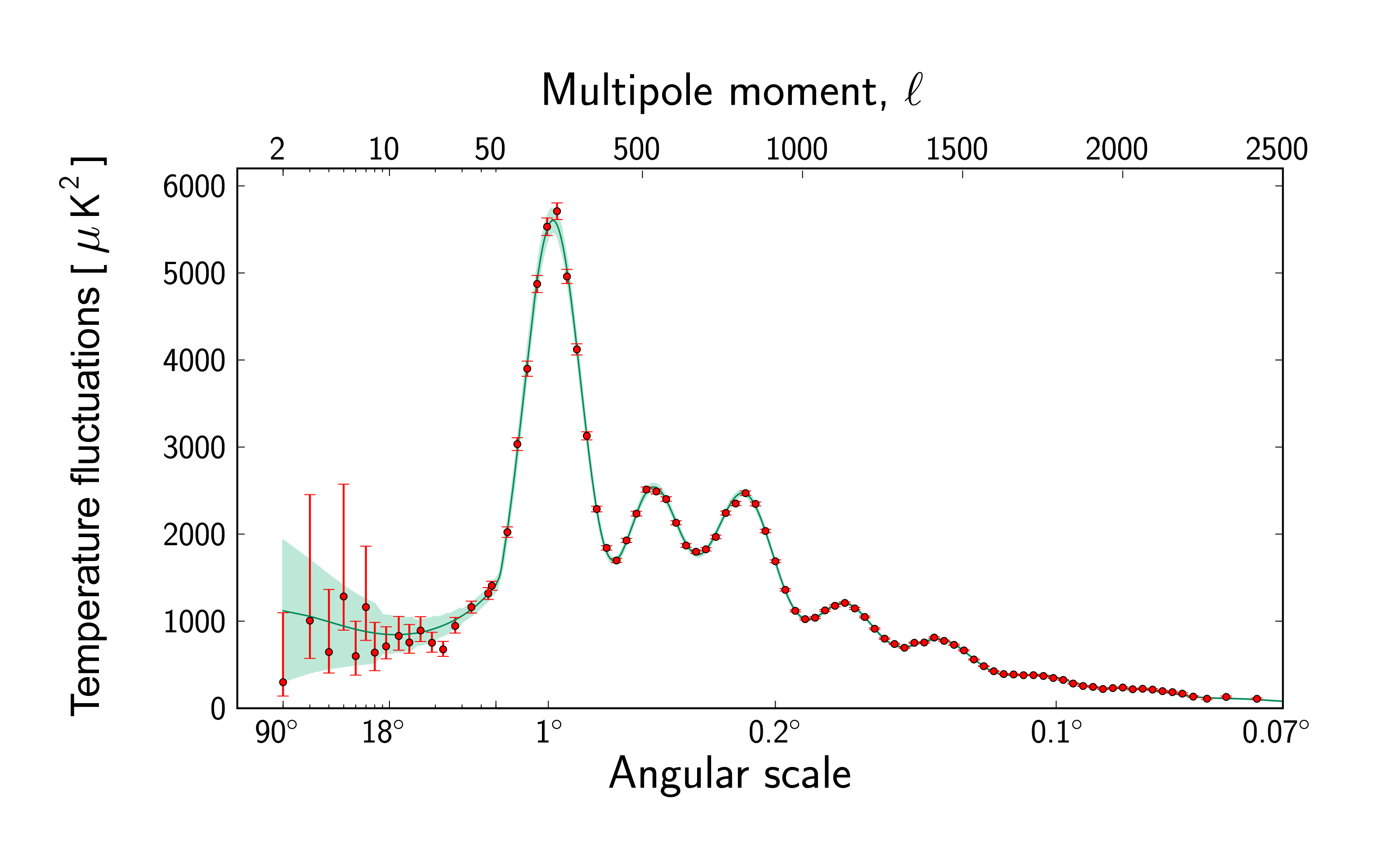}
\caption{\small\label{spectrum}The power spectrum measured by Planck, showing the fluctuations in temperature at a~range of size scales on the sky. 
 The three major peaks show the relative contributions of dark energy, baryonic matter, and dark matter\cite{spectrum}.}
\end{center}
\end{figure}

We can derive information about the role of dark matter in the early Universe from relics, among them the most important
 is the cosmic microwave background radiation (CMB). It~originates from the time of recombination, 
when electrons were bounded with protons to form neutral atoms. From that time, photons could freely stream
through the Universe, what is now observed as a radiation with perfect black body spectrum of temperature $2.725$~K. 
The CMB is nearly homogeneous. Although the fluctuations, that mimic the density anisotropies in the plasma before recombination
are of order $10^{-5}$, they can be used to derive baryon and matter distribution in the Universe. The CMB power spectrum in the angular decomposition 
(fig.~\ref{spectrum}) includes number of peaks that come from acoustic oscillations of primordial plasma. These oscillations
were driven by gravitational attraction of matter and the pressure of radiation. Their behaviour was imprinted in the CMB at the moment of recombination.
The odd peaks refer to compression of plasma and the even ones to its rarefaction, while baryon density affects only the latter \cite{nature}. 
Thus one can measure discrepancy between number of baryons and total matter distribution. Fitting the power spectrum to the $\Lambda\text{CMD}$ model 
yields the precise values (\ref{values}) of dark matter, dark energy and baryon density.

To summarise, there is clear evidence for existence of dark matter on vast range of cosmic scales. Many cosmological observations are in the remarkable
mutual agreement and create a self-consistence model called the "Concordance Cosmology", where dark matter is the essential ingredient.
Nevertheless the cosmology alone cannot explain the role of dark matter in the~particle physics context.

\chapter{Methods of dark matter detection}
There are basically three different methods that may be employed to identify dark matter particles, all of them relies on the assumption
that DM particles interact with the SM also through other forces than gravity. We hope to produce the DM particles at colliders, because they allow
to control initial conditions  and obtain precise results, however no signals of dark matter were found at LHC. Beside collider experiments
there are also numerous efforts to detect dark matter directly in laboratories or indirectly looking for products of its annihilation.

The goal of the direct detection experiments is to measure recoiled energies of nuclei scattered by the DM particles with unknown mass $m_\chi$.
It can be estimated that DM particles present in our galaxy halo should flow through the Earth with a flux
 $10^5(100\;\text{GeV}/m_\chi)\text{cm}^{-2}\text{s}^{-1}$.
The~obtained rate and energy of the nuclear recoils can be used to reconstruct properties of dark matter \cite{bertone}. 
There are two leading technologies in the current detectors: cryogenic detectors measure the heat, that is produced when particle 
interacts with solid state or superfluid~$^3$He absorbers at temperature below $100$~mK and scintillation detectors, which
use light emitted during a collision of dark matter with liquefied noble gas, such as argon or xenon. 
All types of detectors are placed deep underground to ensure screening from cosmic rays and the environmental radioactivity.

There was an impressive improvement in the detectors precision over the last years, mostly in the mass range below $100\;\text{GeV}$, where the direct detection
methods complements collider searches. For the weakly interacting particles and typical galactic velocities 
of order~$v/c\sim 10^{-3}$ scattering of the nuclei should take place at or below $10^{-8}$~pb, 
which is in the sensitivity range of current detectors. In the recent years there were numerous claims of dark matter detection.
The most promising from the DAMA/LIBRA experiment, which announced the observation of the annual modulation
of the nuclear recoil rate, a possible effect that comes from the motion of the Earth around the Sun. In spite of the fact that the signal is strong, its interpretation
is problematic, because it was not confirmed by any other experiments. An excess of events, connected with interaction of dark matter particles,
 was also reported by COGENT, CRESST, EGRET and others  \cite{bergstrom}. Nevertheless direct detection methods are still inconclusive, 
because of large amount of uncertainty, e.g.~from nucleon matrix elements or the nuclear form factors.

Indirect detection aims at finding the secondary particles produced by annihilation or decay of dark matter particles, 
mainly in the center of our galaxy. Recently most studies concentrate
on $\gamma$-rays, because they travel rather unabsorbed through galaxy in straight lines and indicate the point, where 
annihilation of dark matter took place. Due to this properties one can separate the energy distribution of the signal from the astrophysical background
\cite{bergstrom}. An interesting case is the annihilation into two photons considered in \cite{bergstrom1988}. Due to conservation of energy and~momentum 
non-relativistically moving particle will annihilate into photons with energy nearly equal to its~rest mass $E_\gamma = m_\chi$. 
Moreover motion of a particle and the Doppler effect results in broadening of a spectral line by only~$10^{-3}$. Unfortunately such processes
are still difficult to observe, because of the annihilation rates suppression (interaction between photons and neutral dark matter must be loop-induced) and
a resolution of the current detectors, which lead to smearing of the signal and its disappearance in the background. Indirect searches are not so competitive 
as direct detection, but there are many preliminary results. There is excess of positrons observed by the satellite PAMELA \cite{adriani} and gamma-line found by FERMI-LAT
\cite{fermi-lat}. The latter signal is difficult to explain by astrophysical sources and corresponds, if confirmed to DM particle with mass greater than 
$130$~GeV.

\chapter{Standard Model with $ \boldsymbol{\nu_R}$}
\section{Description of $SM\nu_R$}

The Standard Model with right-handed neutrinos $\nu_R$ consists of the following matter fields:

\begin{table*}[htb]\centering
\renewcommand{\arraystretch}{1.4}
\begin{tabular}{|l|cccccc|c|}\hline
 & \mc{6}{c|}{fermions}& scalars\\\hline
field& $l^j_{Lp}$ & \mc{1}{c}{$e_{Rp}$} & \mc{1}{c}{$\nu_{Rp}$} & \mc{1}{c}{$q^{\alpha j}_{Lp}$} 
& \mc{1}{c}{$u^\alpha_{Rp}$} & \mc{1}{c|}{$d^\alpha_{Rp}$} & $\varphi^j$\\\hline
hypercharge $Y$ & $-\frac{1}{2}$ & \mc{1}{c}{$-1$} & \mc{1}{c}{$0$} & \mc{1}{c}{$\frac{1}{6}$} &
 \mc{1}{c}{$\frac{2}{3}$} & \mc{1}{c|}{$-\frac{1}{3}$} & $\frac{1}{2}$\\
\hline
\end{tabular}
\end{table*}
\noindent Weak isospin, colour and generation indices are denoted by $j=1,2$, $\alpha=1,2,3$ and $p=1,2,3$ respectively.
$\tilde\varphi$ is defined as $\tilde{\varphi}_i\equiv\varepsilon_{ij}(\varphi^{j})^*$
. Tensors $\varepsilon_{ij}$ and $\varepsilon_{\mu\nu\rho\sigma}$ are totally antisymmetric with $\varepsilon_{12}=+1$, $\varepsilon_{0123}=+1$.
Dual tensor to $X_{\mu\nu}$ is defined as 
$\tilde{X}_{\mu\nu}=\frac{1}{2}\varepsilon_{\mu\nu\rho\sigma} X^{\rho\sigma}$. Symbol $(\sim)$ over ${X}$ denotes $X$ or $\tilde{X}$.
\\

The SM Lagrangian with gauge-singlet neutrino $\nu_R$ is
\begin{equation}
\begin{split}
\mathcal{L}^{(4)}_{SM} = &-\frac{1}{4}G^A_{\mu\nu}G^{A\mu\nu} -\frac{1}{4}W^I_{\mu\nu}W^{I\mu\nu} -\frac{1}{4}B_{\mu\nu}B^{\mu\nu}
+(D_\mu \varphi)^\dagger(D^\mu \varphi) + m^2 \varphi^\dagger\varphi - \frac{1}{2}\lambda(\varphi^\dagger\varphi)^2\\
&+i(\bar{l}\slashed{D} l + \bar{\nu}_R\slashed{D} \nu_R+\bar{e}\slashed{D}e+\bar{q}\slashed{D} q+\bar{u}\slashed{D} u+\bar{d}\slashed{D} d)\\
&-(\bar{l}\Gamma_\nu \nu_R\tilde{\varphi}+ \bar{l}\Gamma_e e\varphi + \bar{q}\Gamma_u u \tilde{\varphi} + \bar{q}\Gamma_d d\varphi + h.c.)\\
&-(\nu^T_R  C m_M \nu_R + h. c. )
\label{smlag}
\end{split}
\end{equation}
$\Gamma_{\nu,e,u,d}$ and Majorana mass $m_M$ are matrices in a corresponding generation space. 
Sign convention for covariant derivative is exemplified by
\begin{equation}
 (D_\mu q)^{\alpha j} = \left[(\partial_\mu + i g' Y_q B_\mu)\delta^{\alpha\beta}\delta^{jk} + ig_s T^{A\alpha\beta}G^{A}_\mu\delta^{jk}
 + i g S^{Ijk}W^I_\mu\delta^{\alpha\beta}\right]q^{\beta k}
\end{equation}
$T^A=\frac{1}{2}\lambda^A$ are SU(3) generators with Gell-Mann matrices $\lambda^A$ and $S^I=\frac{1}{2}\tau^I$ are SU(2) generators with Pauli
matrices $\tau^I$. It is useful to define Hermitian derivative term

\begin{equation}
 i\varphi^\dagger \overleftrightarrow{D}_\mu \varphi \equiv i \varphi^\dagger D_\mu \varphi - i (D_\mu \varphi)^\dagger \varphi.
\end{equation}


Gauge field strength tensors and their covariant derivative are 
\begin{equation}
\begin{split}
&G^A_{\mu\nu}=\partial_\mu G^A_\nu - \partial_\nu G^A_\mu - g_s f^{ABC}  G^B_\mu  G^C_\nu,\;\;\;\;
(D_\rho G_{\mu\nu})^A = \partial_\rho G^A_{\mu\nu} - g_s f^{ABC} G^B_\rho  G^C_{\mu\nu} \\
&W^I_{\mu\nu}=\partial_\mu W^I_\nu - \partial_\nu W^i_\mu - g \varepsilon^{IJK}  W^J_\mu  W^K_\nu,\;\;\; 
(D_\rho W_{\mu\nu})^I = \partial_\rho W^I_{\mu\nu} - g \varepsilon^{IJK} W^J_\rho  W^K_{\mu\nu} \\
&B_{\mu\nu}=\partial_\mu B_\nu - \partial_\nu B_\mu,\;\;\;\;\;\;\;\;\;\;\;\;\;\;\;\;\;\;\;\;\;\;\;\;\;\;\;\;\;\;\;\;\;
D_\rho B_{\mu\nu}= \partial_\rho B_{\mu\nu}\\
\end{split}
\end{equation}

Equations of motion read as follows:
\begin{equation}
\begin{split}
&(D_\mu D^\mu\varphi)^j=m^2\varphi^j - 
\lambda (\varphi^\dagger\varphi)\varphi^j-\varepsilon_{jk}\bar{l}^k\Gamma_\nu \nu_R 
-\bar{e}\Gamma^\dagger_e l^j -
\varepsilon	_{jk}\bar{q}^k\Gamma_u u - \bar{d}\Gamma^\dagger_d q^j,\\
&(D^\rho G_{\rho\mu})^A=g_s(\bar{q}\gamma_\mu T^Aq + \bar{u}\gamma_\mu T^A u + \bar{d}\gamma_\mu T^A d),\\
&(D^\rho W_{\rho\mu})^I=\frac{g}{2}(\varphi^\dagger i \overleftrightarrow{D}^I_\mu\varphi+ \bar{l}\gamma_\mu \tau^I l +\bar{q}\gamma_\mu \tau^I q),\\
&\partial^\rho B_{\rho\mu} = g'Y_\varphi \varphi^\dagger i \overleftrightarrow{D}_\mu\varphi 
+ g' \sum_{\psi\in\{l,e,q,u,d\}} Y_\psi\bar{\psi}\gamma_\mu\psi,\\
\end{split}
\end{equation}
\begin{equation}
\begin{split}
&i\slashed{D}l=\Gamma_\nu \nu_R \tilde{\varphi} + \Gamma_e e \varphi,\\
&i\slashed{\partial}\nu_R=\Gamma^\dagger_\nu \tilde{\varphi}^\dagger l + m_{M}\nu^C_R, \\\
&i\slashed{D}e=\Gamma^\dagger_e\varphi^\dagger l,\\
&i\slashed{D}q=\Gamma_u u \tilde{\varphi} + \Gamma_d d \varphi,\\
&i\slashed{D}u=\Gamma^\dagger_u \tilde{\varphi}^\dagger q,\\
&i\slashed{D}d=\Gamma^\dagger_d \varphi^\dagger q.
\end{split}
\label{eom}
\end{equation}



\section{Gauge singlet operators in $SM\nu_R$}
The aim of this section is to determine the whole set of gauge singlet operators that consists of $SM\nu_R$ fields. 
The operators are ordered by their dimension and divided into 14 classes: $\boxed{\psi}$, $\boxed{\psi X}$, $\boxed{\psi \varphi}$,
$\boxed{\psi D}$, $\boxed{\psi X \varphi}$, $\boxed{\psi X D}$, $\boxed{\psi X \varphi D}$, $\boxed{\psi \varphi D}$ $\boxed{X}$,
$\boxed{X \varphi}$, $\boxed{X D}$, $\boxed{X \varphi D}$, $\boxed{\varphi}$, $\boxed{\varphi D}$. Each $\psi$, $X$, $\varphi$ or $D$ in a box indicates 
that the class contains operators with respectively at least one: fermion, vector field tensor, scalar or derivative. Operators with 
antisymmetrized covariant derivatives that form a field tensor could appear in arbitrary class with $D$, 
but it is assumed that they are included only in those containing $X$. Therefore class $\boxed{D}$ can be omitted, 
because it consist of derivatives, which need a field to act upon.
%
%
Covariant derivatives satisfy the Leibniz rule and when they act on an object from the singlet 
representation, they are identical to ordinary partial derivatives.
Therefore expanding total derivatives of gauge singlet field operators
(class of such operators is denoted by $\boxed{TD}$ ) all ordinary partial derivatives in the Leibniz rule  can be replaced by covariant ones.
The results of this section are summarised in tab. \ref{tablesm}.

%



\vspace{0.2cm}
DIM 1

No singlets.\\

DIM 1.5

Every operator of non-integer dimension must contain a fermionic field.

\begin{itemize}
 \item $\boxed{\Psi}$

The only fermionic field which is the singlet under all gauge transformations is
\begin{equation}
 \nu_{Rp}.
\end{equation}
\end{itemize}


DIM 2

\begin{itemize}
 \item $\boxed{X}$

\begin{equation}
\overset{(\sim)}{B}_{\mu\nu}.
 \end{equation}

 \item $\boxed{\varphi}$

\begin{equation}
\varphi^\dagger\varphi
 \end{equation}

\end{itemize}

Other classes are empty and therefore omitted.\\

DIM 2.5

\begin{itemize}
\item $\boxed{\psi \varphi }$

Hypercharge U(1) symmetry forces $\varphi$ and $l$, and due to $SU(2)$ symmetry there is only one singlet operator\\
\begin{equation}
\bar{l}\tilde{\varphi}
 \end{equation}

 \item $\boxed{\psi D}$

\begin{equation}
 \partial_\mu \nu_{Rp}
\end{equation} 

\end{itemize}

DIM 3

\begin{itemize}
 \item $\boxed{\psi}$

Each of the $SM\nu_R$ fermions has different absolute value of hypercharge, therefore both fermions in each operator must be the same.
In the basis of 16 Dirac matrices $\gamma^\mu$ and $\gamma^\mu\gamma^5$ change chirality and $1$, $\gamma_5$ and $\sigma_{\mu\nu}$ preserve it. 
Therefore for two Weyl spinors with the same chirality, we have three possible fermionic currents.
\begin{equation}
 \bar{\psi}_p\gamma^\mu\psi_q,\;\;\;\;\;\psi\in\{l,\nu_R,e,q,u,d\}
\end{equation}
\begin{equation}
 \nu^T_{Rp} C\nu_{Rq},\;\;\; \nu^T_{Rp} C\sigma_{\mu\nu}\nu_{Rq}
\end{equation}
The first operator is a gauge singlet for every $SM\nu_R$ fermion, but the currents with particle-antiparticle conjugation $C$ are singlets only for
$\nu_{Rp}$. There are also gauge singlet operators in other Lorentz representations, e.g. $\psi\bar{\psi}$ or $\nu_R\nu_R$, relevant for four-fermion interaction.
Nonetheless, it can be shown that the currents above cover all possibilities in forming four-fermion operators (Appendix \ref{four}).

\item $\boxed{X \varphi}$

No singlets.

 \item $\boxed{X D}$

\begin{equation}
 \partial_\mu \overset{(\sim)}{B}_{\nu\rho}
\end{equation}
\begin{equation}
 \partial^\mu B_{\mu\nu}
\end{equation}
Contraction of Lorentz indices for dual tensor gives an operator that vanish due to the Bianchi identity~$\boxed{BI}$
\begin{equation}
 \frac{1}{2}\varepsilon_{\sigma\rho\mu\nu}\partial^\sigma B^{\mu\nu}=0.
\end{equation}

\item $\boxed{\varphi}$

No singlets.

 \item $\boxed{\varphi D}$

\begin{equation}
 \partial_\mu(\varphi^\dagger\varphi),\;\;\;i\varphi^\dagger \overleftrightarrow{D}_\mu\varphi
\end{equation}
These two operators are Hermitian, first is the sum, and the second is the difference times~$i$ of operators $(D_\mu\varphi)^\dagger\varphi$
and $\varphi^\dagger D_\mu\varphi$

\end{itemize}

DIM 3.5

These operators must contain exactly one fermion. 
\begin{itemize}
%
 \item $\boxed{\psi X}$

\begin{equation}
 \nu_{Rp} \overset{(\sim)}{B}_{\mu\nu}
\end{equation}
 \item $\boxed{\psi \varphi}$

$Y$ allows for $\varphi^\dagger\varphi\nu_{Rp}$ and vanishing operator $\varphi^i\varepsilon_{ij}\varphi^j e=0$.
\begin{equation}
 \varphi^\dagger\varphi\nu_{Rp}
\end{equation}

\item $\boxed{\psi D}$

\begin{equation}
 \partial_\mu \partial_\nu \nu_{Rp}
\end{equation}
\begin{equation}
\Box\nu_{Rp}
\end{equation}

\item $\boxed{\psi \varphi D}$

As for DIM 2.5, due to hypercharge, the allowed operators must contain $l$ and $\varphi$.
%
%
%
\begin{equation}
 \partial_\mu (\bar{l}\tilde{\varphi}), \;\;\;(D_\mu \bar{l}) \tilde{\varphi}
\end{equation}
Operator $\bar{l}D_\mu \tilde{\varphi}$ is the difference of the two above.
\end{itemize}

DIM 4\\

\textsc{Lorentz scalars}
\begin{itemize}
 \item $\boxed{\psi\varphi}$

In this class $SU(3)_C\times SU(2)_L\times U(1)_Y$ symmetry allows only Yukawa interaction terms from $SM\nu_R$~Lagrangian.
\begin{equation}
 \bar{l}\nu_{Rp}\tilde{\varphi}
\end{equation}
\begin{equation}
 \bar{l}e{\varphi}
\end{equation}
\begin{equation}
 \bar{q}u\tilde{\varphi}
\end{equation}
\begin{equation}
 \bar{q}d\varphi
\end{equation}

 \item $\boxed{\psi D}$

Each fermion has different hypercharge, therefore we have only
\begin{equation}
 \bar{\psi}_p\slashed{D}\psi_q,\;\;\;\;\;\psi\in\{l,\nu_R,e,q,u,d\}
\end{equation}
\begin{equation}
 (D_\mu\bar{\psi}_p)\gamma^\mu\psi_q,\;\;\;\;\;\psi\in\{l,\nu_R,e,q,u,d\}
\end{equation}
Total derivative is redundant, because $\partial_\mu (\bar{\psi}_p\gamma^\mu\psi_q) = (D_\mu \bar{\psi}_p) \gamma^\mu \psi_q + \bar{\psi}_p\slashed{D}\psi_q$.

 \item $\boxed{X}$

\begin{equation}
 \overset{(\sim)}X_{\mu\nu}X^{\mu\nu},\;\;\; X_{\mu\nu} \in \{B_{\mu\nu}, W^I_{\mu\nu}, G^A_{\mu\nu}\}
\end{equation}

 \item $\boxed{XD}$

Such operators contain two derivatives and field tensor. The only contraction leading to scalar is $D_\mu D_\nu X^{\mu\nu}$.
Antisymmetrization of derivatives leads to the operators from previous class.

 \item $\boxed{\varphi}$

\begin{equation}
 (\varphi^\dagger\varphi)^2
\end{equation}

 \item $\boxed{\varphi D}$

\begin{equation}
 (D_\mu\varphi)^\dagger (D^\mu\varphi)
\end{equation}
\begin{equation}
 \varphi^\dagger D_{\mu}D^\mu \varphi
\end{equation}
\begin{equation}
 (D_\mu D^\mu\varphi)^\dagger\varphi
\end{equation}

As in the case of fermions, total derivative is redundant, because $\Box (\varphi^\dagger \varphi) = 
\varphi^\dagger D_\mu D^\mu \varphi + (D_\mu D^\mu \varphi)^\dagger \varphi + D^\mu \varphi^\dagger D_\mu \varphi$.
\end{itemize}

We note that $SM\nu_R$ Lagrangian (\ref{smlag}) contains all Lorentz scalars with the exception of total derivatives 
$\tilde{X}_{\mu\nu}X^{\mu\nu}$ and derivative terms $ (D_\mu\bar{\psi})\gamma^\mu\psi$, $\varphi^\dagger D_{\mu}D^\mu \varphi$, $(D_\mu D^\mu\varphi)^\dagger\varphi$,
which are equivalent to standard kinetic terms up to total derivatives
\begin{equation}
\begin{split}
&\tilde{G}^A_{\mu\nu}G^{A\mu\nu}=4\varepsilon^{\mu\nu\rho\sigma}\partial_\mu\left(G^A_\nu\partial_\rho G^A_\sigma-
\frac{1}{3}g_s f^{ABC}G^A_\nu G^B_\rho G^C_\sigma \right),\\
&\tilde{W}^I_{\mu\nu}W^{I\mu\nu}=4\varepsilon^{\mu\nu\rho\sigma}\partial_\mu\left(W^I_\nu\partial_\rho W^I_\sigma-
\frac{1}{3}g \varepsilon^{IJK}W^I_\nu W^J_\rho W^K_\sigma \right),\\
&\tilde{B}_{\mu\nu}B^{\mu\nu}=4\varepsilon^{\mu\nu\rho\sigma}\partial_\mu\left(B_\nu\partial_\rho B_\sigma\right),\\
&(D_\mu\bar{\psi})\gamma^\mu\psi=\partial_\mu (\bar{\psi}\gamma^\mu\psi) - \bar{\psi}\slashed{D}\psi,\\
& (D_\mu D^\mu \varphi)^\dagger \varphi = \partial_\mu [(D^\mu\varphi)^\dagger\varphi] - (D^\mu\varphi)^\dagger D_\mu\varphi,\\
& \varphi^\dagger D_\mu D^\mu \varphi = \partial_\mu (\varphi^\dagger D^\mu \varphi) - (D^\mu\varphi)^\dagger D_\mu\varphi.
\label{smscalars}
\end{split}
\end{equation}



\break
\textsc{Lorentz vectors}

\begin{itemize}

 \item $\boxed{\psi\varphi}$

The only singlets with a scalar and fermions are Yukawa terms in $\mathcal{L}^{(4)}_{SM}$.
They contain fermions with different helicities and do not allow the $\gamma_\mu$ except of the term with $\nu_R$ because its conjugation 
does not spoil the gauge invariance
\begin{equation}
 \nu^T_{Rp} C \gamma^\mu l \varepsilon \varphi.
\label{psiphi}
\end{equation}

 \item $\boxed{\psi D}$

The only possibility is to put $D$ into DIM3 operators which are lorentzian scalars or tensors: 
\begin{equation}
 \nu^T_{Rp}  C \partial_\mu \nu_{Rq}
\end{equation}
$(\partial_\mu\nu^T_{Rq}) C \nu_{Rp}$ is equal to the operator above.
\begin{equation}
 \nu^T_{Rp} C  \gamma_{\mu}\slashed{\partial} \nu_{Rq}
\end{equation}
Contraction with tensors can be reduced to the operators above by the formula (\ref{dsigma}) from appendix.
\begin{equation}
\begin{split}
&\nu^T_{Rp} C   \sigma_{\mu\nu}\partial^\mu \nu_{Rq} \overset{\ref{dsigma}}{=}
\nu^T_{Rp} C i(\partial_\nu-\gamma_\nu \slashed{\partial}) \nu_{Rq} \\
\end{split}
\end{equation}



 \item $\boxed{X}$

  No vectors.

 \item $\boxed{X\varphi}$

  No vectors.
 \item $\boxed{XD}$

  No vectors.
 \item $\boxed{X\varphi D}$

Lorentz vector composed of a field tensor must contain one derivative and one scalar field, but there are no such singlets in DIM 4.  

 \item $\boxed{\varphi D}$

To compose a vector, number of derivatives must be odd, but there are no singlets with odd number of $\varphi$
\end{itemize}

\textsc{Lorentz tensors}
\begin{itemize}

\item $\boxed{\psi \varphi}$

Gauge symmetries allow only Yukawa terms with $\sigma^{\mu\nu}$
\begin{equation}
\bar{l}\sigma^{\mu\nu} \nu_{Rp}\tilde{\varphi},\;\;\;
\bar{l}\sigma^{\mu\nu} e\varphi,\;\;\;\bar{q}\sigma^{\mu\nu} u \tilde{\varphi},\;\;\;\bar{q}\sigma^{\mu\nu} d \varphi 
\end{equation}

\item $\boxed{\psi D}$

Both fermions must be the same, because of $Y$. 
\begin{equation}
 \bar{\psi}_pD_\mu\gamma_\nu\psi_q,\;\;\;\partial_\mu(\bar{\psi}_p\gamma_\nu\psi_q),\;\;\;\;\;\psi\in \{l,\nu_R,e,q,u,d\}
\end{equation}

$(D_\mu \bar{\psi}_p)\gamma_\nu\psi_q$ is the difference of the two above.

 \item $\boxed{X}$

\begin{equation}
 \overset{(\sim)}{G^{A\rho}_{\mu}}G_\rho^{A\nu},\;\;\;
\overset{(\sim)}{W^{I\rho}_{\mu}}W_\rho^{I\nu},
\overset{(\sim)}{{B_{\mu}}^{\rho}} {B_\rho}^{\nu}
\end{equation}

\item $\boxed{X\varphi}$

\begin{equation}
 \varphi^\dagger \overset{(\sim)}{W}_{\mu\nu}\varphi,\;\;\;\varphi^\dagger \overset{(\sim)}{B}_{\mu\nu}\varphi
\end{equation}

\item $\boxed{X D}$

It is composed of two derivatives and $B$ field tensor. 
\begin{equation}
 \Box \overset{(\sim)}{B}_{\mu\nu},\;\;\;\partial_\mu \partial^\rho B_{\rho\nu}
\end{equation}
Contraction of derivative and dual tensor vanishes due to the Bianchi identity.

\item $\boxed{X\varphi D}$

No singlets
\item $\boxed{\varphi D}$

There are three possibilities: $\varphi^\dagger D_\mu D_\nu \varphi$, $(D_\mu D_\nu \varphi)^\dagger\varphi$ and
$(D_\mu \varphi)^\dagger D_\nu \varphi$. They can be combined as in DIM3 to more convenient basis:
\begin{equation}
\begin{split}
\partial_\mu\partial_\nu (\varphi^\dagger\varphi)=&
(D_\mu D_\nu \varphi)^\dagger\varphi + \varphi^\dagger D_\mu D_\nu \varphi+
(D_\mu \varphi)^\dagger D_\nu \varphi + (D_\nu \varphi)^\dagger D_\mu \varphi,\\
\partial_\mu(i\varphi^\dagger \overleftrightarrow{D}_\nu \varphi)=&
i(D_\mu \varphi)^\dagger D_\nu \varphi - i (D_\nu \varphi)^\dagger D_\mu \varphi + i \varphi^\dagger D_\mu D_\nu \varphi-
i(D_\mu D_\nu \varphi)^\dagger\varphi ,\\
(D_\mu \varphi)^\dagger D_\nu \varphi.
\end{split}
\end{equation}

\end{itemize}

\chapter{Dark matter operators}

The aim of this section is to find a list of all\footnote{We omit Lorentz vectors and symmetric tensors of dimension 4, 
because the $SM\nu_R$ does not contain respective operators up to dimension 2. 
Therefore they are redundant to find the effective Lagrangian for $SM\nu_R \times DM$ interactions up to dimension~6, which we obtain in this thesis.}  
operators up to dimension 4, that consist of Dark Matter fields: a real scalar $\Phi$, 
left and right chiral fermions $\Psi_L,\Psi_R$ and a vector field $V_\mu$. Throughout the whole thesis all these fields are assumed to be
singlets of the SM gauge group and in this part we do not presuppose the existence of any symmetry stabilising DM particles. 
The results are summarised in the table \ref{tabledm}.

\section{DM operators with scalar $\Phi$ and fermions $\Psi_L,\Psi_R$.}

Each $\Psi$ without a subscript denotes $\Psi_L$ or $\Psi^c_R$. Operators are as in the SM case ordered by dimension and classes: $\boxed{\Psi}$, 
$\boxed{\Psi\Phi}$, $\boxed{\Psi D}$, $\boxed{\Psi \Phi D}$, $\boxed{\Phi}$, $\boxed{\Phi D}$. Symbols $\Psi$, $\Phi$, $D$ in boxes
indicates that operators of a given class contain respectively fermions, scalars or derivatives.\\


DIM 1

\begin{itemize}
 \item $\boxed{\Phi}$
\begin{equation}
 \Phi
\end{equation}
\end{itemize}
%
%


DIM 1.5

\begin{itemize}
 \item $\boxed{\Psi}$
\begin{equation}
 \Psi
\end{equation}
\end{itemize}

DIM 2

\begin{itemize}
%

\item $\boxed{\Phi}$
\begin{equation}
 \Phi^2
\end{equation}

\item $\boxed{\Phi D}$
\begin{equation}
 \partial_\mu \Phi
\end{equation}

\end{itemize}

DIM 2.5

\begin{itemize}

 \item $\boxed{\Psi \Phi}$
\begin{equation}
 \Psi\Phi
\end{equation}

 \item $\boxed{\Psi D}$
\begin{equation}
 \partial_\mu\Psi
\end{equation}

\end{itemize}




DIM 3

\begin{itemize}
 \item $\boxed{\Psi}$

There are 3 possible currents:
\begin{equation}
\Psi^T C \Psi,\;\;\;\bar{\Psi}\gamma_\mu\Psi,\;\;\;\Psi^T C \sigma_{\mu\nu} \Psi
\end{equation}


%
%
%
%
%
%
%
%
%
%
%
%
%

 \item $\boxed{\Phi}$
\begin{equation}
 \Phi^3
\end{equation}

 \item $\boxed{\Phi D}$
\begin{equation}
 \partial_\mu(\Phi^2),\;\;\; \Box \Phi,\;\;\;\partial_\mu \partial_\nu \Phi
\end{equation}
Operator $\Phi \partial_\mu\Phi$ is redundant, because it equals $\frac{1}{2}\partial_\mu(\Phi^2)$.

%
%

%

\end{itemize}

DIM 3.5
\begin{itemize}

 \item $\boxed{\Psi \Phi}$
\begin{equation}
 \Psi \Phi^2
\end{equation}

 \item $\boxed{\Psi D}$
\begin{equation}
 \Box \Psi
\end{equation}
\begin{equation}
 \partial_\mu\partial_\nu \Psi
\end{equation}

%

%
%
%

 \item $\boxed{\Psi \Phi D}$

\begin{equation}
 \Psi \partial_\mu\Phi,\;\;\;\Phi\partial_\mu\Psi
\end{equation}


%
\end{itemize}





DIM 4\\

In the $SM\nu_R$ there are only scalars and antisymmetric tensors of DIM 1 and DIM 2, therefore it is enough to find respective DM operators to 
construct effective operators for $SM\nu_R\times DM$ interactions up to dimension 6. \\

\textsc{Lorentz scalars} \\


\begin{itemize}
%

\item $\boxed{\Psi \Phi}$

\begin{equation}
 \Phi\Psi^T C\Psi
\end{equation}

\item $\boxed{\Psi D}$

\begin{equation}
(\partial_\mu\bar{\Psi})\gamma^\mu\Psi\;\;\; \bar{\Psi}\slashed{\partial}\Psi
\end{equation}
$\partial_\mu (\bar{\Psi}\gamma^\mu \Psi)$ is their combination.
%
%
%

\item $\boxed{\Phi}$

\begin{equation}
 \Phi^4
\end{equation}

\item $\boxed{\Phi D}$

\begin{equation}
 \Phi\Box\Phi,\;\;\;\partial_\mu \Phi \partial^\mu \Phi
\end{equation}
Operator $\Box(\Phi^2)$ is redundant, because it is equal to
$2 \Phi\Box \Phi + 2 \partial_\mu \Phi \partial^\mu \Phi$


\end{itemize}

\textsc{Lorentz antisymmetric tensors}

\begin{itemize}

\item $\boxed{\Psi\Phi}$

\begin{equation}
 \Psi^T C\sigma^{\mu\nu}\Psi\Phi
\end{equation}
		
\item $\boxed{\Psi D}$

 \begin{equation}
  \partial_\mu(\bar{\Psi}\gamma_\nu\Psi),\;\;\; \bar{\Psi}\partial_\mu\gamma_\nu\Psi
 \end{equation}

%
%
%
%
%
%
%
%
%

\item $\boxed{\Phi D}$

 All tensors in this class are symmetric.

\end{itemize}

%
%
%
%
%
%
%
%
%
%

\section{DM operators with an additional vector field $V_\mu$}
\subsection{Mass generation for DM vector field}

A massive vector field can be described with the use of Proca Lagrangian (\ref{proca}), but the mass term spoils the gauge invariance and 
leads to a term in propagator that gives rise to quadratic divergences at high energies,
this is the feature that we will try to avoid.
\begin{equation}
\begin{split}
 &\mathcal{L}_P=-\frac{1}{4}V_{\mu\nu}V^{\mu\nu} + \frac{1}{2}m_V^2 V_\mu V^\mu\\
 &V_{\mu\nu}=\partial_\mu V_\nu - \partial_\nu V_\mu
\label{proca}
\end{split}
\end{equation}

In all known renormalizable theories, massive vector field is treated as a gauge field. 
One of them is the Stuckelberg mechanism given by Lagrangian $\mathcal{L}_S$. 
The introduction of real scalar field $\sigma$ restores gauge symmetry
and makes the theory manifestly renormalizable \cite{Ruegg}, \cite{Hees}.
\begin{equation}
\begin{split}
&\mathcal{L}_S= -\frac{1}{4}V_{\mu\nu}V^{\mu\nu} + \frac{1}{2}(\partial_\mu \sigma - m_V V_\mu)(\partial^\mu \sigma - m_V V^\mu)\\
&V_\mu \rightarrow V'_\mu= V_\mu +\partial_\mu\chi\\
&\sigma \rightarrow \sigma' = \sigma + m_V\chi 
\label{stuckel}
\end{split}
\end{equation}
Field $\sigma$ can be eliminated from the model by choosing $\chi=-\frac{\sigma}{m_V}$. In this gauge condition Stuckelberg Lagrangian 
becomes the same as in the Proca theory. The Stuckelberg mechanism allows the interaction of massive vector field with conserved currents.
Field $V_\mu$ can couple to scalars and fermions through a covariant derivative
\begin{equation}
 D_\mu \pi = (\partial_\mu - ig V_\mu)\pi.
\label{stuckscal}
\end{equation}
The gauge invariance is still present if $\pi$ transforms as
\begin{equation}
 \pi\rightarrow\pi'= e^{+ig\chi}\pi.
\end{equation}
There is also possible interaction of $V_\mu$ with another vector field $X$ through the following kinetic term
\begin{equation}
 V_{\mu\nu}X^{\mu\nu}
\end{equation}
Such interactions are usually removed from Lagrangians by diagonalization of kinetic terms of all vector fields in the model.

Another way to make a vector field massive is the Higgs mechanism. It uses complex scalar field $\phi$ with nonzero vacuum expectation value
$\langle \phi\rangle=\frac{f}{\sqrt{2}}$. Due to mechanism of spontaneous symmetry breaking, field $V_\mu$ acquires mass $m_V=gf$.
\begin{equation}
\begin{split}
 &\mathcal{L}_H= -\frac{1}{4}V_{\mu\nu}V^{\mu\nu} + (D_\mu\phi)^\dagger D^\mu \phi - \lambda\left(\phi^\dagger\phi - \frac{f^2}{2}\right)^2\\
&D_\mu \phi = (\partial_\mu - i g V_\mu)\phi.\\
&V_\mu \rightarrow V'_\mu - \partial_\mu \chi\\
&\phi \rightarrow \phi' e^{ig\chi}. 
\label{higgs}
\end{split}
\end{equation}
In contrast to the Stuckelberg mechanism the scalar field cannot be completely eliminated by the gauge transformation. 
We can write the complex scalar field as $\phi=(\frac{f}{\sqrt{2}}+\frac{h}{\sqrt{2}})e^{i\theta/f}$. 
In terms of these real scalar fields Higgs the  Lagrangian looks like
\begin{equation}
\begin{split}
 \mathcal{L}_H =& -\frac{1}{4}V_{\mu\nu}V^{\mu\nu} +
 \frac{1}{2}\left( \partial_\mu h-i(1+h/f)\partial_\mu\theta+ig(f+h)V_\mu\right)\left( \partial^\mu h+i(1+h/f)\partial^\mu\theta-ig(f+h)V^\mu\right)\\
&-\frac{\lambda}{2}(4h^2f^2+4h^3 f + h^4)
\end{split}
\end{equation}
The field $h$ has a mass $m_h=2\sqrt{\lambda}f$. We can remove field $\theta$ from the theory choosing unitary gauge 
$\phi \rightarrow \phi e^{-i\theta/f}$, $V_\mu\rightarrow V_\mu+\frac{1}{f}\partial_\mu\theta$. The interactions of $h$ and $V_\mu$ in this 
gauge are:
\begin{equation}
 2 g^2 f h V_\mu V^\mu, \;\;\; g^2h^2 V_\mu V^\mu,\;\;\;-2\lambda h^3 f,\;\;\; -\frac{\lambda}{2}h^4.
\label{higgsint}
\end{equation}

The Stuckelberg mechanism can be seen as a limit of the Higgs mechanism. 
If $f\rightarrow \infty$ and $g\rightarrow 0$, but their product is preserved, then
vector mass $m_V$ remains constant. Scalar field $h$ decouples from $\theta$ and $V_\mu$ and its mass tends to infinity $m_h\rightarrow\infty$. 
The Higgs Lagrangian $\mathcal{L}_H$ becomes identical as in the Stuckelberg mechanism with vector mass $m_V=fg$.

\begin{equation}
 \mathcal{L}^{\text{lim}}_H=-\frac{1}{4}V_{\mu\nu}V^{\mu\nu} + \frac{1}{2}(\partial_\mu \theta - fg V_\mu)(\partial^\mu \theta - fg V^\mu)+\mathcal{L}_h,
\end{equation}
where $\mathcal{L}_h$ is Lagrangian of the decoupled field $h$. The field $h$ can be also decoupled within the effective field theory framework. 
If we assume that $V_\mu$ is a light particle with mass of the order~$v$, which is SM Higgs VEV, 
we can put $h$ into the heavy particles spectrum (in the theory with high-energy scale $\Lambda$).  
Scalar mass is $m_h=2\sqrt{\lambda}f  \sim \sqrt{\lambda} \frac{v}{g}$, therefore for $\lambda \sim 1$ it is enough to set coupling $g \sim \frac{v}{\Lambda}$.

\subsection{Operators within the Stuckelberg mechanism}

When the mass of the vector field is generated by the Stuckelberg mechanism(\ref{stuckel}) the gauge invariance requires
that the vector field appears in the model as $V_{\mu\nu}$, $\tilde{V}_{\mu\nu}$ or $\mathcal{V}_\mu=\partial_\mu \sigma - m_V V_\mu$. 
It is assumed, that $V_\mu$ is not a gauge vector for additional $U(1)$ symmetry of dark matter or standard model fields.
Note that, the field $\mathcal{V}_\mu$ has dimension 2.\\

DIM 1

No gauge invariant operators with $V_\mu$.\\

DIM 1.5

No gauge invariant operators with $V_\mu$.\\

DIM 2

\begin{equation}
 \mathcal{V}_\mu
\end{equation}
\begin{equation}
  \overset{(\sim)}{V}_{\mu\nu}
\end{equation}

DIM 2.5

No gauge invariant operators with $V_\mu$.\\

DIM 3

Field tensor  $\overset{(\sim)}{V}_{\mu\nu}$ can be combined with the scalar field $\Phi$ or with a derivative:
\begin{equation}
 \overset{(\sim)}{V}_{\mu\nu} \Phi
\end{equation}
\begin{equation}
 \partial^\mu{V}_{\mu\nu}
\end{equation}
\begin{equation}
 \partial_\rho\overset{(\sim)}{V}_{\mu\nu}
\end{equation}
Dual tensor contracted with derivative vanishes due to the Bianchi identity. The other possibility is to use the field $\mathcal{V}_\mu$:
\begin{equation}
 \mathcal{V}_\mu\Phi
\end{equation}
\begin{equation}
 \partial_\mu\mathcal{V}_\nu
\end{equation}
\begin{equation}
 \partial_\mu\mathcal{V}^\mu
\end{equation}

DIM 3.5

\begin{equation}
 \mathcal{V}_\mu\Psi
\end{equation}
\begin{equation}
 \overset{(\sim)}{V}_{\mu\nu} \Psi
\end{equation}

\break
DIM 4

As for the operators without the vector field, only Lorentz scalars and symmetric tensors are relevant.\\

\textsc{Lorentz scalars}

\begin{equation}
 \overset{(\sim)}{V}_{\mu\nu}V^{\mu\nu}
\end{equation}
\begin{equation}
 \mathcal{V}_\mu\mathcal{V}^{\mu}
\end{equation}
\begin{equation}
 \partial_\mu \mathcal{V}^{\mu} \Phi
\end{equation}
\begin{equation}
 \mathcal{V}_{\mu}  \partial^\mu \Phi
\end{equation}
\vspace{0cm}

\textsc{Lorentz antisymmetric tensors}

${V}_{\mu\rho}V^{\rho\nu}$ and $\tilde{V}_{\mu\rho}V^{\rho\nu}$ are symmetric, the latter due to (\ref{symten}).
\begin{equation}
 \Box{V}_{\mu\nu}
\end{equation}
\begin{equation}
  \partial_\mu\partial^\rho {V}_{\rho\nu}
\end{equation}
\begin{equation}
 \partial_\mu \mathcal{V}_{\nu}\Phi
\end{equation}
\begin{equation}
\mathcal{V}_{\nu} \partial_\mu \Phi
\end{equation}
\begin{equation}
 \overset{(\sim)}{V}_{\mu\nu}\Phi^2
\label{phiphiV}
\end{equation}

\subsection{Operators within the Higgs mechanism.}

Gauge invariant quantities containing a vector field in the model with the Higgs mechanism (\ref{higgs}) are built from $V_{\mu\nu}$, $\tilde{V}_{\mu\nu}$
and covariant derivatives of complex scalar field $D_\mu\phi$.  It is assumed that dark matter and standard model fields
are singlets of the Higgs U(1) symmetry. Operators built of $V_{\mu\nu}$ and $\tilde{V}_{\mu\nu}$ 
only are the same as in Stuckelberg case. Operators with $\phi$ appear at DIM 3, because they must contain $\phi^*$ to ensure gauge invariance
and at least one covariant derivative that contains the vector field.\\\\


%
%
%
%
%
%
%
%
%
%

%
%


DIM 3

\begin{equation}
 \phi^* D_\mu \phi
\end{equation}

DIM 3.5

No operators with $\phi$.\\


DIM 4

\textsc{Lorentz scalars}


\begin{equation}
 (D_\mu\phi)^* D^\mu\phi
\end{equation}
\begin{equation}
 \phi^*D_\mu D^\mu \phi,\;\;\;(D_\mu D^\mu \phi)^*\phi
\end{equation}

\textsc{Lorentz antisymmetric tensors}

%
%
%

\begin{equation}
   (D_\mu\phi)^* D_\nu\phi
\end{equation}
Antisymmetrized operators operators $\phi^*D_\mu D_\nu \phi$ and $(D_\mu D_\nu \phi)^*\phi$ are equivalent to \ref{phiphiV}.

\begin{table}
\begin{center}
\te
\small
\begin{tabular}{|l|l|l|l|}
\hline
DIM & no space-time indices & one space-time index $\mu$ & more space-time indices $\mu,\nu,\rho$\\
\hline
1 & - & - & -\\\hline
1.5 & $\nu_{Rp}$ & - & -\\\hline
2 & $\varphi^\dagger\varphi$ & - & $\overset{(\sim)}{B}_{\mu\nu}$\\\hline
2.5 & $\bar{l}\tilde{\varphi}$ & $\partial_\mu \nu_{Rp}$ & -\\\hline
3 & $\nu^T_{Rp} C \nu_{Rq}$ & $\bar{\psi}_p\gamma_\mu \psi_q$, $i\varphi^\dagger\overleftrightarrow{D}_\mu\varphi$ 
& $\nu^T_{Rp} C \sigma_{\mu\nu} \nu_{Rq}$, $\partial_\rho\overset{(\sim)}{B}_{\mu\nu}$\\
& & $\partial_\mu(\varphi^\dagger\varphi)$, $\partial^\mu B_{\mu\nu}$ & \\\hline
3.5 & $\varphi^\dagger\varphi\nu_{Rp}$, $\Box \nu_{Rp}$ 
& $\partial_\mu(\bar{l}\tilde{\varphi})$, $(D_\mu \bar{l})\tilde{\varphi}$ 
& $\partial_\mu\partial_\nu \nu_{Rp}$, $\nu_{Rp} \overset{(\sim)}{B}_{\mu\nu}$\\\hline
4 & $(D_{\mu}\varphi)^\dagger D^\mu\varphi$, $\phi^4$, $\psi_p\slashed{D}\psi_q$, & 
 $\partial_\mu (\nu^T_{Rp} C\nu_{Rq})$, $ \nu^T_{Rp}  C \partial_\mu \nu_{Rq} $, &
$\overset{(\sim)}{X}_{\mu\rho}{X_\nu}^\rho$, $\Box\overset{(\sim)}{B}_{\mu\nu}$,  
 $\varphi^\dagger \overset{(\sim)}{W}_{\mu\nu}\varphi$, $\varphi^\dagger \overset{(\sim)}{B}_{\mu\nu}\varphi $,\\
& $\bar{l}\nu_{Rp}\tilde{\varphi}$, $\bar{l}e{\varphi}$, $\bar{q}u\tilde{\varphi}$, $\bar{q}d\varphi$,
& $\nu^T_{Rp} C  \gamma_{\mu}\slashed{\partial} \nu_{Rq}$, $(\slashed{\partial}\nu^T_{Rp}) C   \gamma_{\mu} \nu_{Rq} $, &
$\partial_\mu\partial_\nu (\varphi^\dagger\varphi)$, $ \partial_\mu(i\varphi^\dagger \overleftrightarrow{D}_\nu \varphi)$,
$(D_\mu\varphi)^\dagger D_\nu \varphi$,
\\
& $\overset{(\sim)}{X}_{\mu\nu}X^{\mu\nu}$, $(D_\mu \bar{\psi}_p)\gamma^\mu \psi_q$,   & $\nu^T_{Rp} C \gamma^\mu l \varepsilon \varphi$ &
$\bar{\psi}_p D_\mu\gamma_\nu\psi_q$, $\partial_\mu(\bar{\psi}_p\gamma_\nu\psi_q)$, $\partial_\mu \partial^\rho B_{\rho\nu}$\\
& $\varphi^\dagger D_\mu D^\mu \varphi$, $(D_\mu D^\mu \varphi)^\dagger \varphi$ & &$\bar{l}\sigma^{\mu\nu} \nu_{Rp} \tilde{\varphi}$, $\bar{l}\sigma^{\mu\nu} e\varphi$ ,
 $\bar{q}\sigma^{\mu\nu} u \tilde{\varphi}$, $\bar{q}\sigma^{\mu\nu} d\varphi $ 
\\
\hline

\end{tabular}
\caption{\small\label{tablesm}SM operators that are singlets of $SU(3)_C\times SU(2)_L\times U(1)_Y$ in different Lorentz group representations. 
$X_{\mu\nu}$ stands for $B_{\mu\nu}$, $W^I_{\mu\nu}$ or $G^A_{\mu\nu}$, $\psi\in \{l,\nu_R,e,q,u,d\}$.}
\end{center}

\begin{center}
\te
\small
\begin{tabular}{|l|l|l|l|}\hline
DIM & no space-time indices & one space-time index $\mu$ & more space-time indices $\mu,\nu,\rho$\\\hline
1 & $\Phi$ & -
 & -\\\hline
1.5 & $\Psi_{L,R}$ & - & - \\\hline 
2 & $\Phi^2$, 
& $\partial_\mu \Phi$, {\color{red}$\mathcal{V}_\mu$}
& 
$\overset{(\sim)}{V}_{\mu\nu}$
\\\hline
2.5 & $\Psi_{L,R}\Phi$ & 
$\partial_\mu\Psi_{L,R}$ 
 & - \\\hline
3 & $\Phi^3$, $\Box \Phi$, 
 $\Psi^T C \Psi$,  {\color{red}$\partial_\mu \mathcal{V}^\mu$}
& $\partial_\mu (\Phi)^2$, $\bar\Psi \gamma_\mu\Psi$, {\color{blue}$\phi^* D_\mu \phi$}
 
& $\partial_\mu\partial_\nu \Phi$,
 $\Psi^T C \sigma_{\mu\nu} \Psi $,

\\
& 
& $\partial^\mu V_{\mu\nu}$, {\color{red}$\mathcal{V}_\mu\Phi$}
& $\Phi\overset{(\sim)}{V}_{\mu\nu}$, {\color{red}$\partial_\mu \mathcal{V}_\nu$},
$\partial_\mu \overset{(\sim)}{V}_{\nu\rho}$\\\hline

3.5 & $\Psi_{L,R}\Phi^2$, $\Box \Psi$, 
& $\Psi_{L,R}\partial_\mu\Phi$, 
$\Phi\partial_\mu\Psi_{L,R}$, {\color{red}$\Psi\mathcal{V}_\mu$}
&
$\partial_\mu\partial_\nu\Psi$, $\Psi\overset{(\sim)}{V}_{\mu\nu}$
\\\hline
4 
& 
$\Phi\Psi^T C \Psi$, $\bar{\Psi}\slashed{\partial}\Psi$, $(\partial_\mu \bar{\Psi})\gamma^\mu \Psi$,
& 
& $\Psi^T C \sigma_{\mu\nu} \Psi \Phi$, $\partial_\mu(\bar{\Psi}\gamma_\nu\Psi)$, 

\\
&
$\Phi^4$, $\partial_\mu\Phi\partial^\mu\Phi$,
$\Phi \Box \Phi$, 
&

&
$ \bar{\Psi}\partial_\mu\gamma_\nu\Psi$,  $\partial_\mu\partial^\rho\overset{(\sim)}{V}_{\rho\mu}$, $\overset{(\sim)}{V}_{\mu\nu}\Phi^2$, 

\\
& 
$\overset{(\sim)}{V}_{\mu\nu} V^{\mu\nu}$, {\color{red}$\mathcal{V}_\mu\mathcal{V}^\mu$}, {\color{red}$\partial_\mu \mathcal{V}^{\mu}\Phi$},
{\color{red}$\mathcal{V}^{\mu}\partial_\mu\Phi$}
&
& 
{\color{red}$\partial_\mu\mathcal{V}_{\nu}\Phi$}, {\color{red}$\mathcal{V}_{\nu}\partial_\mu\Phi$}, 
\\
& {\color{blue}$(D_\mu\phi)^*D^\mu\phi$} ,  {\color{blue}$\phi^*D_\mu D^\mu \phi$}, {\color{blue}$(D_\mu D^\mu \phi)^*\phi$}
&
&  {\color{blue}$(D_\mu \phi)^* D_\nu \phi$}
\\\hline
%

\end{tabular}
\caption{\small\label{tabledm}DM operators with scalar $\Phi$, $\Psi_L$ and $\Psi_R$ and $V_\mu$. Each $\Psi$ denotes $\Psi_L$ or $\Psi^c_R$.
Operators with $\mathcal{V}$ (highlighted with red colour) appear only, when mass of $V_\mu$
is generated by the Stuckelberg mechanism and operators with $\phi$ (blue colour) are present only in the model with the Higgs mechanism. Vector operators 
and symmetric tensors in dimension 4 are not
enlisted, because we are interested in the $SM\nu_R\times DM$ operators up to dimension 6 and the $SM\nu_R$ does not contain vector operators or symmetric tensors with dimension less
or equal 2.
}
\end{center}
\end{table}


\chapter{Renormalizable models of $\boldsymbol{SM\nu_R \times DM}$ interactions.}

The aim of this section is to determine all possible interactions between  the $SM\nu_R$ and
 the $DM$ sector with $\Phi$, $\Psi_L$, $\Psi_R$ and $V_\mu$ 
which are renormalizable, therefore dimension of operators is constrained to be $\leq 4$. We assume that all $DM$ fields are massive.
As before $\Psi$ denotes left-chiral dark fermion or charge conjugation of its right-chiral counterpart
\begin{equation}
 \Psi \in \{\Psi_L, \Psi^c_R\}.
\end{equation}

Lorentz and gauge invariant operators of DM fields with  $DIM\leq 4$  are enlisted in first column of tab.~\ref{tabledm}. 
Many of them can be eliminated from the general Lagrangian: dimension one operator $\Phi$ can be removed by simple translation of this field
by constant $\Phi\rightarrow\Phi+c$, $\Box\Phi$ and $\partial_\mu\mathcal{V}^\mu$ are total derivatives, $(\partial_\mu \bar{\Psi})\gamma^\mu \Psi$ 
is up to total derivative equal to standard kinetic term $\bar\Psi\slashed{\partial}\Psi$, similarly $\Phi\Box\Phi$ is redundant,
$\tilde{V}_{\mu\nu}V^{\mu\nu}=4\varepsilon^{\mu\nu\rho\sigma}\partial_\mu(V_\nu\partial_\rho V_\sigma)$ is also a total derivative.
It should be noted that the operator $\mathcal{V}^\mu\partial_\mu\Phi$, which appears in the Stuckelberg scenario can be removed as follows 
\begin{equation}
\begin{split}
&-\frac{1}{4}V_{\mu\nu}V^{\mu\nu} + \frac{1}{2}(\partial_\mu \sigma - m_V V_\mu)(\partial^\mu \sigma - m_V V^\mu)
+\alpha(\partial^\mu \sigma - m_V V^\mu)\partial_\mu\Phi+\frac{1}{2}\partial_\mu\Phi\partial^\mu\Phi+V(\Phi)=\\
&-\frac{1}{4}V_{\mu\nu}V^{\mu\nu} + \frac{1}{2}(\partial_\mu \sigma + \alpha\partial_\mu\Phi  - m_V V_\mu)
(\partial^\mu \sigma + \alpha\partial^\mu\Phi - m_V V^\mu)+\frac{1}{2}(1-\alpha^2)\partial_\mu\Phi\partial^\mu\Phi + V(\Phi),
\end{split}
\end{equation}
then one can redefine $\sigma$, such that $\sigma\rightarrow \sigma - \alpha \Phi$ and rescale $\Phi\rightarrow \Phi/\sqrt{1-\alpha^2}$.
Eventually the unwanted term disappears. The last redundant operator $\Phi\partial_\mu\mathcal{V}^\mu$ is equal up to total derivative to the previous operator. 
Operators with $\phi$ are omitted, because $\phi$ is treated as a heavy field, not a DM particle. \\

\begin{itemize}
 \item DIM 2
\begin{equation}
 \Phi^2
\end{equation}
 \item DIM 3
\begin{equation}
 \Phi^3,\;\;\;\Psi^T C \Psi 
\end{equation}
 \item DIM 4
\begin{equation}
 \Phi\Psi^T C \Psi,\;\;\; \bar\Psi\slashed{\partial}\Psi,\;\;\;\Phi^4,\;\;\;\partial_\mu\Phi\partial^\mu\Phi,\;\;\;
V_{\mu\nu}V^{\mu\nu},\;\;\;\mathcal{V}_\mu\mathcal{V}^\mu
\end{equation}	
\end{itemize}



A similar list of all $SM\nu_R \times DM$ interactions can be determined using both tables \ref{tablesm} and \ref{tabledm}:

\begin{itemize}
 \item DIM 3

\begin{equation}
 \varphi^\dagger\varphi\Phi
\end{equation}

 \item DIM 4

\begin{equation}
\begin{split}
&\bar{\nu}_{Rp}\Psi\Phi,\;\;\;\varphi^\dagger\varphi\Phi^2,\;\;\;\bar{l}\Psi^c\tilde{\varphi},\;\;\;\nu^T_{Rp} C \nu_{Rq} \Phi,\;\;\;
\overset{(\sim)}{B}_{\mu\nu}V^{\mu\nu}
\end{split}
\end{equation}
\end{itemize}

We can choose operators from the lists above to build models of the $SM\nu_R$ interacting with the DM sector.
The models can be defined imposing symmetries that forbid some operators. Transformation properties 
of the DM candidates are adjusted to ensure their stability.  It is useful to classify those models 
with respect to transformation properties of the $SM\nu_R$ fields, which could either be singlets 
or transform non-trivially under those symmetries. The two options are discussed below.
We assume that at least one of the dark fields must be stable.

\begin{enumerate}
 \item $SM\nu_R$ fields are singlets of $DM$ fields symmetries.

In the model with only scalar field $\Phi$ there is just one $SM\nu_R \times DM$ operator that fulfils our assumptions
\begin{equation}
 \varphi^\dagger\varphi\Phi^2,
\end{equation}
because $\varphi^\dagger\varphi\Phi$ would lead to decay of $\Phi$ into $SM$ fields. Then $\Phi^4$ is the only self-interaction, that ensures stability.  
This model can be easily realised by 
adopting $Z_2$ symmetry for $\Phi$. If dark sector contains also fermions then the $SM\nu_R \times DM$ interactions may also include the following
operators 
\begin{equation}
\bar\nu_{Rp} \Psi \Phi
\end{equation}
since $\Psi \Phi$ can form a singlet of a stabilising symmetry. Note that if $\Phi$ is the lightest
particle transforming under the symmetry, then it remains stable.  
Then $\Phi\Psi_{L}^T C \Psi_{L}$, $\Phi\Psi_{R}^T C \Psi_{R}$ and $\Phi\bar{\Psi}_L\Psi_R$ might be 
responsible for DM self interaction. 
Such model can be realised by $Z_2$ or $Z_2 \times Z_2$ symmetries.

In the presence of a vector field $SM\times DM$ interaction may include $\overset{(\sim)}{B}_{\mu\nu}V^{\mu\nu}$.

Summarising we find the following cases:
\begin{itemize}
\item Symmetry $Z_2$:  $\Phi\to -\Phi$
\begin{equation}
\Phi^4,\;\;\;\varphi^\dagger\varphi \Phi^2
\end{equation}
This model has been extensively discussed in the literature \cite{Drozd2012}.

 \item Symmetry $Z_2$:  $\Phi\to -\Phi$ and $\Psi_L\to -\Psi_L$ (or $\Psi_R\to-\Psi_R$)

For $\Psi_L\to -\Psi_L$ allowed operators are
\begin{equation}
 \bar{\Psi}_L\nu_{Rp}\Phi,\;\;\;\bar{\Psi}_L\Psi_R\Phi,\;\;\;\Phi^4,\;\;\;\varphi^\dagger\varphi \Phi^2,
\end{equation}
while for $\Psi_R\to-\Psi_R$ 
\begin{equation}
 \Psi^T_R C \nu_{Rp}\Phi,\;\;\;\bar{\Psi}_L\Psi_R\Phi,\;\;\;\Phi^4,\;\;\;\varphi^\dagger\varphi \Phi^2,
\end{equation}
 \item Symmetry $Z_2\times Z_2$: $\Psi_L\sim (-,+)$, $\Psi_R\sim (+,-)$, $\Phi\sim (-,-)$ 
\begin{equation}
 \bar{\Psi}_L\Psi_R\Phi,\;\;\;\Phi^4,\;\;\;\varphi^\dagger\varphi \Phi^2
\end{equation}
In this case, there are two $Z_2$ symmetries imposed, therefore two components of the
dark sector could be stable. This model was discussed in details in \cite{twocomp}.
\item Symmetry $Z_2$:  $\Phi\to -\Phi$
\begin{equation}
\overset{(\sim)}{B}_{\mu\nu}V^{\mu\nu},\;\;\;\ \Phi^4,\;\;\;\varphi^\dagger\varphi \Phi^2
\end{equation}
If the dark sector contained fields charged under the (Stuckelberg or Higgs) gauge transformation, then the vector field $V_\mu$ could
interact with them in the standard manner through covariant derivative.

\end{itemize}

 \item $SM\nu_R$ transforms non-trivially under $DM$ fields symmetries.
 
The only interaction term from $SM\nu_R$ Lagrangian that can be removed without destroying its correct experimental predictions is the
Yukawa interaction of $\nu_R$. If we assume that $\nu_R$ transforms non-trivially under $DM$ fields symmetries, the only $SM\nu_R \times DM$ operators 
that are allowed by dark symmetries are: 
\begin{equation}
 \bar{\Psi}\nu_{Rp}\Phi,\;\;\;\bar{\Psi}_L\Psi_R\Phi,\;\;\;
 \nu^T_{Rp} C \nu_{Rq} \Phi,\;\;\; \Phi^4,\;\;\;\varphi^\dagger\varphi\Phi^2
\end{equation}
In terms of $Z_2$ symmetries we find the following realisations (for three $\nu_{Rp}$ flavours):
\begin{itemize}
 \item Symmetry $Z_2$:  $\Phi\to -\Phi$ and $\nu_{R3}\to -\nu_{R3}$ 

Note that, without compromising generality, we can arbitrarily choose the 3rd $\nu_R$ to be odd.
Then the following interactions are allowed:
\begin{equation}
 \nu^T_{R3} C \nu_{Rq} \Phi\; {\rm for}\;\; q=1,2,\;\;\; \Phi^4,\;\;\;\varphi^\dagger\varphi\Phi^2
\end{equation}
$\Phi$ or $\nu_{Rp}$ could be stable in this case.
 \item Symmetry $Z_2$:  $\Phi\to -\Phi$, $\Psi_L\to -\Psi_L$ (or $\Psi_R\to -\Psi_R$)
and $\nu_{R3}\to -\nu_{R3}$ 

Then the following interactions are allowed for $\Psi_L\to -\Psi_L$:
\begin{equation}
 \Psi^T_R C \nu_{R3}\Phi,\;\;\; \bar{\Psi}_L\Psi_R\Phi,\;\;\;
 \nu^T_{R3} C \nu_{Rq} \Phi\; {\rm for}\;\; q=1,2,\;\;\; \Phi^4,\;\;\;\varphi^\dagger\varphi\Phi^2
\end{equation}
Similarly for $\Psi_R\to -\Psi_R$. The lightest dark field could be stable in those cases.
 \item Symmetry $Z_2\times Z_2$: $\nu_{Ri}\sim (-,+)$ (e.g. for $i=3$), 
$\nu_{Rq}\sim (+,-)$ (e.g. for $q=1,2$), $\Phi\sim (-,-)$ 
\begin{equation}
 \nu^T_{Rp}C\nu_{Rq}\Phi,\;\;\; \Phi^4,\;\;\; \varphi^\dagger\varphi \Phi^2
\end{equation}
In this case, $\Phi$ and one of $\nu_{Rp}$ components could are stable.
\end{itemize}


Additional DM fermions can be easily included. Note that $\Psi_R$ and $\Psi^c_L$ can be treated as additional flavours of neutrinos $\nu_{Rp}$.


\end{enumerate}

\chapter{Effective operators for $\boldsymbol{SM\nu_R \times DM}$ interactions}

\section{Effective Lagrangians}

The effective field theory is a method to represent the dynamical content of a theory in the low energy limit. 
It appears naturally, when there is a hierarchy of mass scales. To analyse a~physical process
below the high energy scale $\Lambda$, one may skip the heavier fields and take into account their effects by adding to the Lagrangian 
new couplings between the light fields \cite{smdynamics,pokorski}. Such process is called "integrating out'' the fields
and leads to the effective Lagrangian, that consist of a series of higher dimensional operators suppressed by the energy scale $\Lambda$

\begin{equation}
 \mathcal{L}=\mathcal{L}^{(4)}+\frac{1}{\Lambda}\sum_{k}C^{5}_k O_k^{(5)}+\frac{1}{\Lambda^2}\sum_{k}C^{6}_k O_k^{(6)},
\end{equation}
where $\mathcal{L}^{(4)}$ is the standard Lagrangian of the theory with light fields only and operators up to dimension 4, $O_k^{(n)}$~denote dimension-$n$
effective operators and $C^{n}_k$ are the corresponding coupling constants (Wilson coefficients). If we neglect the influence of higher dimensional operators
the~physics at lower energy scale is described by renormalizable part of the Lagrangian. This is the result of
the Appelquist-Carrazone \cite{appelquist} theorem, which 
states that all effects, that depend logarithmically or with higher powers on $\Lambda$, can be absorbed into the renormalizable
Lagrangian by the redefinition of the bare parameters.
 
However the correct version of a high energy theory is not known, we expect one with the new scale e.g. at the Planck level.
The effective field theory is a useful tool to parametrize its effects, because the higher dimensional operators 
are strongly suppressed and the series in the effective Lagrangian can be quickly truncated. It is natural to assume that the physics at lower energies is described by 
higher dimensional operators that obey symmetries of the low energy theory. In the case of the SM the full set of such operators up to dimension
 6 was constructed in~\cite{wyler} and then reexamined \cite{iskrzynski}. 
In this thesis we analyse the effective operators in the extensions of the $SM\nu_R$ with additional fields, that are candidates for dark matter.
\vspace{2cm}

\section{$SM\nu_R \times DM$ operators}

In this section it is assumed that each $DM$ field: $\Phi$, $\Psi_L$, $\Psi_R$ and $V_\mu$ is odd under $Z_2$ symmetry\footnote{
In other words the Lagrangian is invariant under independent $Z_2$ corresponding to each dark field. As a consequence each component of the dark sector
is stable. This assumption could be relaxed allowing for the presence of messenger fields that couple both to the SM and dark sector.}.
Therefore only interactions quadratic in $DM$ fields are allowed. Lagrangian $\mathcal{L}^{(4)}$ up to dimension 4 consists of 3 parts
\begin{equation}
 \mathcal{L}^{(4)}=\mathcal{L}^{(4)}_{SM}+\mathcal{L}^{(4)}_{DM}+\mathcal{L}^{(4)}_{SM\times DM}.
\end{equation}
The first is Lagrangian of the Standard Model with right-chiral neutrinos
\begin{equation}
\begin{split}
\mathcal{L}^{(4)}_{SM} = &-\frac{1}{4}G^A_{\mu\nu}G^{A\mu\nu} -\frac{1}{4}W^I_{\mu\nu}W^{I\mu\nu} -\frac{1}{4}B_{\mu\nu}B^{\mu\nu}
+(D_\mu \varphi)^\dagger(D^\mu \varphi) + m^2 \varphi^\dagger\varphi - \frac{1}{2}\lambda(\varphi^\dagger\varphi)^2\\
&+i(\bar{l}\slashed{D} l+ + \bar{\nu}_R\slashed{D} \nu_R+\bar{e}\slashed{D}e+\bar{q}\slashed{D} q+\bar{u}\slashed{D} u+\bar{d}\slashed{D} d)\\
&-(\bar{l}\Gamma_\nu \nu_R\tilde{\varphi}+ \bar{l}\Gamma_e e\varphi + \bar{q}\Gamma_u u \tilde{\varphi} + \bar{q}\Gamma_d d\varphi + h.c.)\\
&-\frac{1}{2}(\nu^T_R  C m_M \nu_R + h. c. ).
\label{SMLag}
\end{split}
\end{equation}
The second is a Lagrangian that contains $DM$ fields only\footnote{Note that Dirac mass term is forbidden by $Z_2$ symmetries.}
\begin{equation}
\begin{split}
 \mathcal{L}^{(4)}_{DM} =  & \frac{1}{2} \partial_\mu \Phi \partial^\mu \Phi -\frac{1}{2} m^2_\Phi \Phi^2 - \frac{1}{4}\kappa \Phi^4 \\
 & -\frac{1}{4} V^{\mu\nu} V_{\mu\nu} + \frac{1}{2} m_V^2 V_\mu V^\mu \\
 & + i (\bar{\Psi}_L \slashed{\partial} \Psi_L + \bar{\Psi}_R\slashed{\partial}\Psi_R)
- \frac{1}{2}(m_L \Psi^T_L C \Psi_L + m_R \Psi^T_R C \Psi_R + h. c.).
\label{DMLag}
\end{split}
\end{equation}
There is only one operator contributing to  $DM\times SM$ interactions
\begin{equation}
  \mathcal{L}^{(4)}_{SM\times DM}=g_\varphi \varphi^\dagger\varphi \Phi^2.
\end{equation}

Equations of motions derived from the $\mathcal{L}^{(4)}$ for $SM\nu_R$ fields are the same as in the SM with one exception. 
Equations for scalars $\varphi$ and $\Phi$ contain a term, that originates from interaction in $\mathcal{L}^{(4)}_{SM\times DM}$.
Complete list of equations for $DM$ and $SM\nu_R$ fields reads as:

\begin{equation}
 \begin{split}
&(D_\mu D^\mu\varphi)^j=m^2\varphi^j - 
\lambda (\varphi^\dagger\varphi)\varphi^j-\varepsilon_{jk}\bar{l}^k\Gamma_\nu \nu_R 
-\bar{e}\Gamma^\dagger_e l^j -
\varepsilon_{jk}\bar{q}^k\Gamma_u u - \bar{d}\Gamma^\dagger_d q^j + g_\varphi\varphi \Phi^2,\\
&(D^\rho G_{\rho\mu})^A=g_s(\bar{q}\gamma_\mu T^Aq + \bar{u}\gamma_\mu T^A u + \bar{d}\gamma_\mu T^A d),\\
&(D^\rho W_{\rho\mu})^I=\frac{g}{2}(\varphi^\dagger i \overleftrightarrow{D}^I_\mu\varphi+ \bar{l}\gamma_\mu \tau^I l +\bar{q}\gamma_\mu \tau^I q),\\
&\partial^\rho B_{\rho\mu} = g'Y_\varphi \varphi^\dagger i \overleftrightarrow{D}_\mu\varphi 
+ g' \sum_{\psi\in\{l,e,q,u,d} Y_\psi\bar{\psi}\gamma_\mu\psi
\end{split}
\end{equation}
\begin{equation}
\begin{split}
&i\slashed{D}l=\Gamma_\nu \nu_R \tilde{\varphi} + \Gamma_e e \varphi,\\
&i\slashed{\partial}\nu_R=\Gamma^\dagger_\nu \tilde{\varphi}^\dagger l + m_{M}\nu^C_R \\\
&i\slashed{D}e=\Gamma^\dagger_e\varphi^\dagger l,\\
&i\slashed{D}q=\Gamma_u u \tilde{\varphi} + \Gamma_d d \varphi,\\
&i\slashed{D}u=\Gamma^\dagger_u \tilde{\varphi}^\dagger q,\\
&i\slashed{D}d=\Gamma^\dagger_d \varphi^\dagger q\\
\end{split}
\end{equation}
\begin{equation}
\begin{split}
&\partial_\mu \partial^\mu \Phi = -m_\Phi \Phi - \kappa \Phi^3 + 2 g_\varphi \Phi \varphi^\dagger \varphi,\\
&\partial^\mu V_{\mu\nu} = - m_V V_\nu,\;\;\;\;\;\;\;\partial_\mu V^{\mu}=0,\\
&i \slashed{\partial} \Psi_{L,R} = m_{L,R} \Psi^c_{L,R}.
\end{split}
\label{fulleomv}
\end{equation}

Many of the $SM\nu_R\times DM$ operators  are redundant due to the equations of motion. 
If an operator includes the LHS of one of the above equations, then
it can be written as a sum of the operators that consists of the RHS of that equation and the operator, denoted by $\boxed{EOM}$,
which vanishes due to that equation 
of motion. If an equation contains reference to an operator number in box (e.g. $\boxed{\ref{vvPP}}$), 
it means that the box should be replaced by the operator up to Hermitian conjugation with a coefficient, that is irrelevant. The purpose is to express given operator
as linear combination of others, total derivatives $\boxed{TD}$ and $\boxed{EOM}$. Such operators are redundant in effective Lagrangian \cite{Knetter}.
Operators vanishing due to the Bianchi identity are denoted by $\boxed{BI}$. Adopting tab. \ref{tablesm} and \ref{tabledm} one can construct all the $SM\nu_R\times DM$ operators
up to dimension~6. Results are contained in the following sections.\\

\section{$SM\nu_R \times DM$  operators with a scalar field $\Phi$}

\indent 

DIM 4

\begin{itemize}
 \item $2\times 2 $
\begin{equation}
 \varphi^\dagger\varphi\Phi^2
\label{vvPP}
\end{equation}
\end{itemize}

DIM 5

\begin{itemize}
 \item $3\times 2$

\begin{equation}
 \nu^T_{Rp} C \nu_{Rq} \Phi^2
\label{rrPP}
\end{equation}
\end{itemize}

DIM 6

\begin{itemize}
 \item $2\times 4$

\begin{equation}
 \varphi^\dagger\varphi \partial_\mu \Phi \partial^\mu \Phi
\label{vvdPdP}
\end{equation}
\begin{equation}
 \varphi^\dagger\varphi \Phi^4
\label{vvPPPP}
\end{equation}
\begin{equation}
\begin{split}
  \varphi^\dagger\varphi \Phi \Box \Phi &= \boxed{EOM} - \varphi^\dagger\varphi \Phi (m_\Phi^2 \Phi - \lambda_\Phi \Phi^3 -
 2 g_\varphi \varphi^\dagger\varphi\Phi)\\ &= \boxed{EOM} + \boxed{\ref{vvPP}} + \boxed{\ref{vvPPPP}} + \boxed{\ref{vvvvPP}}
\end{split}
\label{vvddPP}
\end{equation}

 \item $3\times 3$

\begin{equation}
\begin{split}
 &\bar{\psi}\gamma_\mu\psi \partial^\mu  (\Phi^2) = \boxed{TD} - \partial_\mu  (\bar{\psi}\gamma_\mu\psi)\Phi^2 = 
\boxed{TD} - (\bar{\psi} \slashed{D}\psi + h. c. )\Phi^2 = \\& \boxed{TD} + \boxed{EOM} + \boxed{\ref{SMPP}} + \boxed{\ref{rrPP}}  
\end{split}
\end{equation}
Operator $\boxed{\ref{rrPP}}$ is relevant only for $\psi=\nu_R$
\begin{equation}
 \partial_\mu (\varphi^\dagger\varphi)\partial^\mu (\Phi^2) = \boxed{TD} - \varphi^\dagger\varphi \Box (\Phi^2) =
\boxed{TD} + \boxed{\ref{vvdPdP}} + \boxed{\ref{vvddPP}} 
\end{equation}
\begin{equation}
\begin{split}
  &\varphi^\dagger\overleftrightarrow{D}_\mu\varphi\partial_\mu (\Phi^2) = 
\boxed{TD} - \partial_\mu (\varphi^\dagger\overleftrightarrow{D}^\mu\varphi) \Phi^2 = 
\boxed{TD} -(\varphi^\dagger D^\mu D_\mu \varphi - h. c.) \Phi^2 = \boxed{TD} + \boxed{EOM} -  \\
& \left[\varphi^{\dagger j}(m^2\varphi^j - 
\lambda (\varphi^\dagger\varphi)\varphi^j-g_\varphi\Phi^2 \varphi^j -\varepsilon_{jk}\bar{l}^k\Gamma_\nu \nu_R 
-\bar{e}\Gamma^\dagger_e l^j -
\varepsilon_{jk}\bar{q}^k\Gamma_u u - \bar{d}\Gamma^\dagger_d q^j - h. c.)\right] \Phi^2=\\
& \boxed{TD} + \boxed{EOM} +\boxed{\ref{vvPP}}+ \boxed{\ref{SMPP}}
\end{split}
\end{equation}
\begin{equation}
\begin{split}
 &\partial_\mu B^{\mu\nu} \partial_\nu (\Phi^2) = \boxed{TD} - \partial_\nu \partial_\mu B^{\mu\nu} \Phi^2 = \boxed{TD}
\end{split}
\end{equation}

 \item $4\times 2$

There are many operators that can be put in one category

\begin{equation}
 O^{(4)}_{SM} \times \Phi^2,
\label{SMPP}
\end{equation}
namely: 
\begin{equation}
 X_{\mu\nu}X^{\mu\nu}\Phi^2,
\end{equation}
\begin{equation}
 (D_\mu \varphi)^\dagger D^\mu\varphi \Phi^2,
\end{equation}
\begin{equation}
 (\varphi^\dagger\varphi)^2\Phi^2,
\label{vvvvPP}
\end{equation}
\begin{equation}
 \bar{l}\nu_{Rp}\tilde{\varphi}\Phi^2
\label{y1}
\end{equation}
\begin{equation}
 \bar{l}e\varphi\Phi^2
\label{y2}
\end{equation}
\begin{equation}
 \bar{q}u\tilde{\varphi}\Phi^2
\label{y3}
\end{equation}
\begin{equation}
 \bar{q}d\varphi\Phi^2
\label{y4}
\end{equation}
Remaining operators contain kinetic terms for $SM\nu_R$ fermions, that can be reduced with equations of motion to the Yukawa interactions and
the Majorana mass term for $\nu_R$ 
\begin{equation}
 i\bar{\psi}\slashed{D}\psi \Phi^2 = \boxed{EOM} + \boxed{\ref{rrPP}}+ \boxed{\ref{SMPP}} .
\end{equation}
$-i(D_\mu \bar{\psi})\gamma^\mu \psi \Phi^2$ is Hermitian conjugation of the operator above.
\begin{equation}
\tilde{X}_{\mu\nu} X^{\mu\nu} \Phi^2
\end{equation}
\begin{equation}
\begin{split}
&\varphi^\dagger D_\mu D^\mu \varphi \Phi^2 = \boxed{EOM} + \\
&\varphi^\dagger \left[m^2\varphi^j - 
\lambda (\varphi^\dagger\varphi)\varphi^j -g_\varphi \Phi^2 \varphi^j -\varepsilon_{jk}\bar{l}^k\Gamma_\nu \nu_R 
-\bar{e}\Gamma^\dagger_e l^j -
\varepsilon_{jk}\bar{q}^k\Gamma_u u - \bar{d}\Gamma^\dagger_d q^j\right]\Phi^2 =\\
&\boxed{EOM}  + \boxed{\ref{vvPP}} +  \boxed{\ref{vvvvPP}} + \boxed{\ref{vvPPPP}} + \boxed{\ref{y1}}+ \boxed{\ref{y2}}+ \boxed{\ref{y3}}+ \boxed{\ref{y4}}
\label{vddvPP}
\end{split}
\end{equation}
$(D_\mu D^\mu \varphi)^\dagger \varphi \Phi^2$ is Hermitian conjugation of \ref{vddvPP}.
\end{itemize}
%
%
%
%
%
%
%

To sum up for DIM 6 the operators are: $\varphi^\dagger\varphi \partial_\mu \Phi \partial^\mu \Phi$, $\varphi^\dagger\varphi \Phi^4$,
$\overset{(\sim)}{X}_{\mu\nu} X^{\mu\nu} \Phi^2$, $(D_\mu \varphi)^\dagger D^\mu\varphi \Phi^2$, 
$(\varphi^\dagger\varphi)\Phi^2$, $\bar{l}\nu_{Rp}\tilde{\varphi}\Phi^2$,
$\bar{l}e\varphi\Phi^2$, $\bar{q}u\tilde{\varphi}\Phi^2$ and $\bar{q}d\varphi\Phi^2$. 

\section{$SM\nu_R \times DM$ operators with fermions: $\Psi_{L}$,$\Psi_R$.}

\indent

DIM 4

No operators.\\

DIM 5

\begin{itemize}
 \item $2\times 3$

\begin{equation}
 \varphi^\dagger\varphi \Psi^T C \Psi
\label{ppSS}
\end{equation}
\begin{equation}
 \overset{(\sim)}{B}_{\mu\nu} \Psi^T C \sigma^{\mu\nu} \Psi
\label{bPsP}
\end{equation}
\end{itemize}

DIM 6

\begin{itemize}
 \item $2\times 4$

\begin{equation}
 \tilde{B}_{\mu\nu} \partial^\mu (\Psi \gamma^\nu \Psi) = \boxed{TD} - (\partial^\mu \tilde{B}_{\mu\nu}) \Psi \gamma^\nu \Psi = \boxed{TD}
\label{DpPgP}
\end{equation}
\begin{equation}
\begin{split}
  &B_{\mu\nu} \partial^\mu (\Psi \gamma^\nu \Psi) = \boxed{TD} - (\partial^\mu B_{\mu\nu}) \Psi \gamma^\nu \Psi = \boxed{TD} + \boxed{EOM} +\\
&-(g'Y_\varphi \varphi^\dagger i \overleftrightarrow{D}_\nu\varphi 
+ g' \sum_{\psi=l,e,q,u,d} Y_\psi\bar{\psi}\gamma_\nu\psi) \Psi \gamma^\nu \Psi = \\
&  \boxed{TD} + \boxed{EOM}+\boxed{\ref{vdvSgS}} +  \boxed{\ref{sgsSgS}}
\label{BpPgP}
\end{split}
\end{equation}
\begin{equation}
 \varphi^\dagger \varphi \bar{\Psi} \slashed{\partial} \Psi = \boxed{EOM} + \boxed{\ref{ppSS}} 
\end{equation}
$\varphi^\dagger \varphi (\partial_\mu\bar{\Psi}) \gamma^\mu \Psi$ is Hermitian conjugation of the operator above.
\begin{equation}
\begin{split}
  \overset{(\sim)}{B^{\mu\nu}}\bar{\Psi} { \gamma_\mu \partial_\nu} \Psi = &\frac{1}{2} \overset{(\sim)}{B^{\mu\nu}}\bar{\Psi}
  (\gamma_\mu \gamma_\nu \slashed{\partial} + \gamma_\mu \slashed{\partial} \gamma_\nu)\Psi =  \frac{1}{2} \overset{(\sim)}{B^{\mu\nu}}\bar{\Psi}
  (\gamma_\mu \gamma_\nu \slashed{\partial}-\slashed{\partial} \gamma_\mu \gamma_\nu)\Psi+
 \overset{(\sim)}{B^{\mu\nu}}\bar{\Psi}\gamma_\nu \partial_\mu \Psi=\\
  &\frac{1}{4} \overset{(\sim)}{B^{\mu\nu}}\bar{\Psi}
  (\gamma_\mu \gamma_\nu \slashed{\partial} - \slashed{\partial} \gamma_\mu \gamma_\nu)\Psi =
 \frac{1}{4} \overset{(\sim)}{B^{\mu\nu}}\bar{\Psi}\gamma_\mu \gamma_\nu \slashed{\partial}\Psi + 
\frac{1}{4}\bar{\Psi}\overleftarrow{\slashed{\partial}}\gamma_\mu \gamma_\nu \Psi   \overset{(\sim)}{B^{\mu\nu}} +\\
&+\frac{1}{4}\bar{\Psi}\gamma_\rho\gamma_\mu\gamma_\nu\Psi \partial^\rho \overset{(\sim)}{B^{\mu\nu}} + \boxed{TD} =\\ 
&\boxed{EOM} + \boxed{TD} + \boxed{\ref{bPsP}} + \boxed{BI} + \boxed{\ref{vdvSgS}} + \boxed{\ref{sgsSgS}},
\label{longest}
\end{split}
\end{equation}
The third equality follows from the fact that last term in the first line is equal to the first with an opposite sign. We also used
the following reduction
\begin{equation}
\begin{split}
 \bar{\Psi}\gamma_\rho\gamma_\mu\gamma_\nu\Psi \partial^\rho \overset{(\sim)}{B^{\mu\nu}} \overset{\ref{ggg}}{=}
 &2\bar{\Psi}\gamma^\nu \Psi \partial^\rho \overset{(\sim)}{B_{\rho\nu}} 
-\bar{\Psi}i\varepsilon_{\rho\mu\nu\sigma}\gamma^\sigma\gamma_5 \Psi \partial^\rho \overset{(\sim)}{B^{\mu\nu}} = \\
=\;\;&\boxed{EOM} + \boxed{BI} + \boxed{\ref{vdvSgS}} + \boxed{\ref{sgsSgS}}. 
\end{split}
\label{gggB}
\end{equation}
Note that in the case of dual tensor the first term vanishes due to the Bianchi identity and the second is reduced by the equation of motion, but
for the standard tensor the situation is reversed.
The reduction to $\boxed{\ref{vdvSgS}}$ and $\boxed{\ref{sgsSgS}}$ is made similarly as in $\boxed{\ref{BpPgP}}$.

 \item $3\times 3$

\begin{equation}
 \nu^T_{Rp} C \nu_{Rq} \Psi^T C \Psi
\end{equation}
\begin{equation}
 \nu^T_{Rp} C \sigma^{\mu\nu} \nu_{Rq} \Psi^T C \sigma_{\mu\nu} \Psi
\end{equation}
These two operators were created by the multiplication of two non-Hermitian operators of DIM 3. 
Therefore we have to include also the respective terms, where one of the~DIM~3 operators is conjugated. Generally, they can have different Wilson
coefficients in the~effective Lagrangian.
\begin{equation}
 \nu^T_{Rp} C \nu_{Rq} \bar{\Psi} C \bar{\Psi}^T
\end{equation}
\begin{equation}
 \nu^T_{Rp} C \sigma^{\mu\nu} \nu_{Rq} \bar{\Psi} \sigma_{\mu\nu} C \bar{\Psi}^T
\end{equation}

Vectorial currents are Hermitian therefore we need only one operator.
\begin{equation}
 \bar{\psi}_p \gamma_\mu \psi_q \bar{\Psi} \gamma^\mu \Psi
\label{sgsSgS}
\end{equation}
\begin{equation}
 i\varphi^\dagger\overleftrightarrow{D}_\mu\varphi\bar\Psi \gamma^\mu \Psi
\label{vdvSgS}
\end{equation}

\begin{equation}
\begin{split}
 \partial_\mu(\varphi^\dagger\varphi)\bar\Psi \gamma^\mu \Psi &= \boxed{TD} - \varphi^\dagger\varphi\partial_\mu(\bar\Psi\gamma^\mu \Psi) =\\
 &=\boxed{TD} - \varphi^\dagger\varphi (\bar\Psi\slashed{\partial} \Psi + h. c.) = \\
&=\boxed{TD} 
- \varphi^\dagger\varphi \bar{\Psi}_{L,R}(m_{L,R} \Psi^c_{L,R}+ h. c.) + \boxed{EOM} = \\
&= \boxed{TD} + \boxed{EOM} + \boxed{\ref{ppSS}}
\end{split}
\end{equation}
\begin{equation}
\begin{split}
 \partial_\mu B^{\mu\nu}\bar\Psi \gamma_\nu \Psi &= \boxed{TD} + \boxed{EOM} +
 \left(g'Y_\varphi \varphi^\dagger i \overleftrightarrow{D}_\mu\varphi 
+ g' \sum_{\psi=l,e,q,u,d} Y_\psi\bar{\psi}\gamma_\mu\psi\right)  \bar\Psi\gamma^\mu \Psi =\\
& =\boxed{TD} + \boxed{EOM} + \boxed{\ref{sgsSgS}} + \boxed{\ref{vdvSgS}}
\end{split}
\end{equation}
\end{itemize}

%
%
%
%
%
%
%
%
%
%
%
%
%
To sum up for DIM 6 the operators are: 
$
 \nu^T_{Rp} C \nu_{Rq} \Psi^T C \Psi
$,
$
\nu^T_{Rp} C \nu_{Rq} \bar{\Psi} C \bar{\Psi}^T
$,
$
 \nu^T_{Rp} C \sigma^{\mu\nu} \nu_{Rq} \Psi^T C \sigma_{\mu\nu} \Psi
$,
$
 \nu^T_{Rp} C \sigma^{\mu\nu} \nu_{Rq}\bar{\Psi} \sigma_{\mu\nu} C \bar{\Psi}^T
$,
$
 \bar\psi_p \gamma_\mu \psi_q \bar\Psi \gamma^\mu \Psi
$,
$
 i\varphi^\dagger\overleftrightarrow{D}_\mu\varphi\bar\Psi \gamma^\mu \Psi	
$.

\section{$SM\nu_R \times DM$ operators with vector field $V_\mu$}

%
%
%

It is assumed that $V_\mu$ is odd under $Z_2$ symmetry to make it stable and that other $DM$ fields are not charged under U(1) symmetry 
with gauge vector $V_\mu$.



\subsection{The Stuckelberg mechanism}


In this model the Lagrangian (\ref{stuckel}) is invariant under $Z_2$ symmetry, when both $\sigma$ and $V_\mu$ are odd with respect to it. 
Equations of motion coming from (\ref{stuckel}) are 
\begin{equation}
\begin{split}
 &\partial_\mu(\partial^\mu\sigma - m_V V^\mu)=\partial_\mu  \mathcal{V}^\mu = 0\\
 &\partial_\mu V^{\mu\nu} =  -m^2_V V^\nu + m_V \partial^\nu \sigma = m_V \mathcal{V}^\nu.
\end{split}
\end{equation}
The Stuckelberg Lagrangian and equations of motion reduce to a part of the Lagrangian (\ref{DMLag}) and equations (\ref{fulleomv}) when a gauge condition with $\sigma=0$
is adopted.\\\\

DIM 6

\begin{itemize}
 \item 
$2\times 4$

\begin{equation}
\varphi^\dagger\varphi \overset{(\sim)}{V}_{\mu\nu}V^{\mu\nu}
\end{equation}
\begin{equation}
 \varphi^\dagger\varphi\mathcal{V}_\mu\mathcal{V}^\mu 
\label{ppVV}
\end{equation}
All these operators have to be suppressed by a factor $\Lambda^{-2}$ in Lagrangian density, which has dimension 4.
It  is  worth to consider closely (\ref{ppVV}) in the $\sigma=0$ gauge
\begin{equation}
  \frac{1}{\Lambda^2} \mathcal{V}_\mu\mathcal{V}^\mu \varphi^\dagger\varphi\rightarrow \frac{m_V^2}{\Lambda^2} V_\mu V^\mu  \varphi^\dagger\varphi,
\end{equation}
Note that higher dimensional operators will be suppressed by more powers of $\Lambda$.

\end{itemize}

%
%
%


%

\subsection{The Higgs mechanism}

Another method to generate vector mass is the Higgs mechanism (\ref{higgs}). It is easy to see that
operators containing $\phi$ and $D_\mu$ are invariant under $Z_2$ for odd $V_\mu$ if $\phi$ transforms into its conjugation\footnote{In this case
$Z_2$ can be seen as a charge conjugation.} 
$\phi\rightarrow \phi^*$ (in other words the phase $\theta$ of $\phi$ is odd, while the absolute value $h$ is even). Eventually
only quadratic terms in $V_\mu$ survive.
Mass $m_V$ is of order of Higgs field VEV $v$ if the gauge coupling constant $g\sim \frac{v}{\Lambda}$ 
and VEV of $\phi$ is of order of heavy scale $\Lambda$. Then Higgs boson appearing in this model can be decoupled in the effective theory.\\

DIM 6
\begin{itemize}
 \item $2 \times 4$
\begin{equation}
 \frac{1}{\Lambda^2}\overset{(\sim)}{V}_{\mu\nu}V^{\mu\nu}\varphi^\dagger\varphi
\end{equation}
\begin{equation}
 \frac{1}{\Lambda^2}(D_\mu \phi)^*D^\mu\phi \varphi^\dagger \varphi
\label{pppp}
\end{equation}
These operators have to be suppressed by a factor
 $\Lambda^{-2}$ in Lagrangian density. When $\phi$ acquires VEV $f=\langle\phi\rangle\sim \Lambda$ and $g\sim v/\Lambda$  (to put Higgs boson 
into heavy sector) then
\begin{equation}
  \frac{1}{\Lambda^2}(D_\mu \phi)^*D^\mu\phi \varphi^\dagger \varphi
 \rightarrow \frac{v^2}{\Lambda^2}  V_\mu V^\mu \varphi^\dagger\varphi = \frac{v^2}{\Lambda^2}  V_\mu V^\mu \varphi^\dagger\varphi,
\end{equation}
so we can conclude that this coupling is also suppressed in the Higgs scenario.

The operator $\phi^*D_\mu D^\mu \phi$ is not invariant under $Z_2$. In order to make it invariant we have to add its conjugation.
\begin{equation}
 (\phi^*D_\mu D^\mu \phi+(D_\mu D^\mu \phi)^*\phi)\varphi^\dagger\varphi = 
\Box(\phi^*\phi)\varphi^\dagger\varphi + \boxed{\ref{pppp}}
\end{equation}
The operator $\Box(\phi^*\phi)$ does not contain vector field, therefore it is irrelevant here.
Similarly $(D_\mu\phi)^*D_\nu\phi + (D_\nu\phi)^*D_\mu\phi$ is invariant under $Z_2$, but this operator is symmetric in its tensorial indices
and vanish after contraction with ${B}_{\mu\nu}$.
\begin{equation}
 \left[(D^\mu\phi)^*D^\nu\phi + (D^\nu\phi)^*D^\mu\phi\right]\overset{(\sim)}{B}_{\mu\nu}=0
\end{equation}
Operator $\phi^* D_\mu \phi$ can be made symmetric under $Z_2$, if we add to it conjugation $(D_\mu\phi)^*\phi$, but
operator $\phi^* D_\mu \phi+(D_\mu\phi)^*\phi=\partial_\mu(\phi^*\phi)$ does not contain vector field $V_\mu$.

\end{itemize}

\begin{table*}[ht]
\centering
\renewcommand{\arraystretch}{1.4}
\begin{tabular}{!{\vrule width 2pt}c!{\vrule width 2pt}c!{\vrule width 2pt}c!{\vrule width 2pt}c|c|c!{\vrule width 2pt}}
\noalign{\hrule height 2pt}
& $1$ & $\Lambda^{-1}$ & \mc{3}{c!{\vrule width 2pt}}{$\;\;\;\;\Lambda^{-2}$} \\\noalign{\hrule height 2pt}
& & & $\Phi$ & $\Psi$ & $V_{\mu}$ \\\hline
\textsc{Tree}:& $\varphi^\dagger\varphi \Phi^2$ &  $\varphi^\dagger\varphi \Psi^T C \Psi$ &  $\varphi^\dagger\varphi \partial_\mu \Phi \partial^\mu \Phi$
& $\nu^T_{Rp} C \nu_{Rq} \Psi^T C \Psi$ & $v^2\varphi^\dagger\varphi V_\mu V^\mu$\\
&& $\nu^T_{Rp} C \nu_{Rq} \Phi^2$ & $(D_\mu \varphi)^\dagger D^\mu \varphi\Phi^2$
& $\nu^T_{Rp} C \sigma^{\mu\nu} \nu_{Rq} \Psi^T C \sigma_{\mu\nu} \Psi$ & \\
&&  & $\varphi^\dagger\varphi \Phi^4$
& $\nu^T_{Rp} C \nu_{Rq} \bar{\Psi} C \bar{\Psi}^T$ &\\
&&  & $(\varphi^\dagger\varphi)^2\Phi^2$
& $\nu^T_{Rp} C \sigma^{\mu\nu} \nu_{Rq} \bar{\Psi} \sigma_{\mu\nu} C \bar{\Psi}^T$ &\\
&& & $\bar{l}\nu_{Rp}\tilde{\varphi}\Phi^2$& $\bar{\psi}_p \gamma_\mu \psi_q \bar\Psi \gamma^\mu \Psi$&\\
&& & $\bar{l}e{\varphi}\Phi^2$ & $ i\varphi^\dagger\overleftrightarrow{D}_\mu\varphi\bar\Psi \gamma^\mu \Psi$&\\
&& & $\bar{q}u\tilde{\varphi}\Phi^2$ &&\\
&& & $\bar{q}d{\varphi}\Phi^2$ &&\\\hline
\textsc{Loop:}&& $\overset{(\sim)}{B}_{\mu\nu} \Psi^T C \sigma^{\mu\nu} \Psi$& & $\overset{(\sim)}{X}_{\mu\nu} X^{\mu\nu} \Phi^2$ &
$\varphi^\dagger\varphi\overset{(\sim)}{V}_{\mu\nu}V^{\mu\nu}$\\
&& & & & $m_V^2\varphi^\dagger\varphi V_\mu V^\mu$\\
\noalign{\hrule height 2pt}

\end{tabular}
\caption{\small\label{tablesmdm}List of all $SM\nu_R\times DM$ operators up to dimension 6, that are suppressed by at most~$\Lambda^{-2}$.
Dark matter sector consists of real scalar $\Phi$, left and right chiral fermions $\Psi_L, \Psi_R$ and vector field $V_\mu$. 
Tree and loop-generated operators are collected in the upper and lower part of the table, respectively.
Operator $\varphi^\dagger\varphi V_\mu V^\mu$ appears in both categories, because within the Higgs mechanism it can be generated at tree-level approximation,
however within the Stuckelberg model it requires a loop.
Note that one~entry in the table may refer to various operators, because $(\sim)$ over $X_{\mu\nu}$ denotes $X_{\mu\nu}$ or~$\tilde{X}_{\mu\nu}$,
 $X_{\mu\nu}$ stands for $B_{\mu\nu}$, $W^I_{\mu\nu}$ or $G^A_{\mu\nu}$, $\psi\in \{l,\nu_{R},e,q,u,d\}$ and $\Psi\in\{\Psi_L,\Psi^c_R\}$.
The bosonic operators are all Hermitian. In case of the operators containing fermions, 
$i\varphi^\dagger\protect\overleftrightarrow{D}_\mu\varphi\bar\Psi \gamma^\mu \Psi$  is Hermitian and conjugation of 
$\bar{\psi}_p \gamma_\mu \psi_q \bar\Psi \gamma^\mu \Psi$ is equivalent to transposition of the generation indices. 
For the remaining operators Hermitian conjugations are not listed explicitly.}
\end{table*}
\section{Generation of effective operators in high-energy theory}
\begin{figure}[b]
\centering
\includegraphics[scale=0.4]{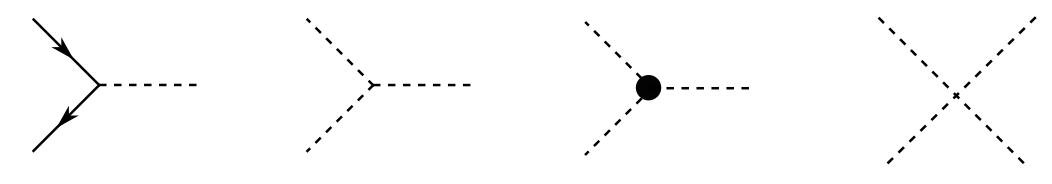}
\caption{\small\label{vertices}List of all possible vertices in underlying field theory. Each dashed line corresponds to a scalar or vector. 
The dot indicates the coupling constant proportional to heavy scale $\Lambda$.}
\end{figure}
\begin{figure}[ht]
\centering
\includegraphics[scale=0.44]{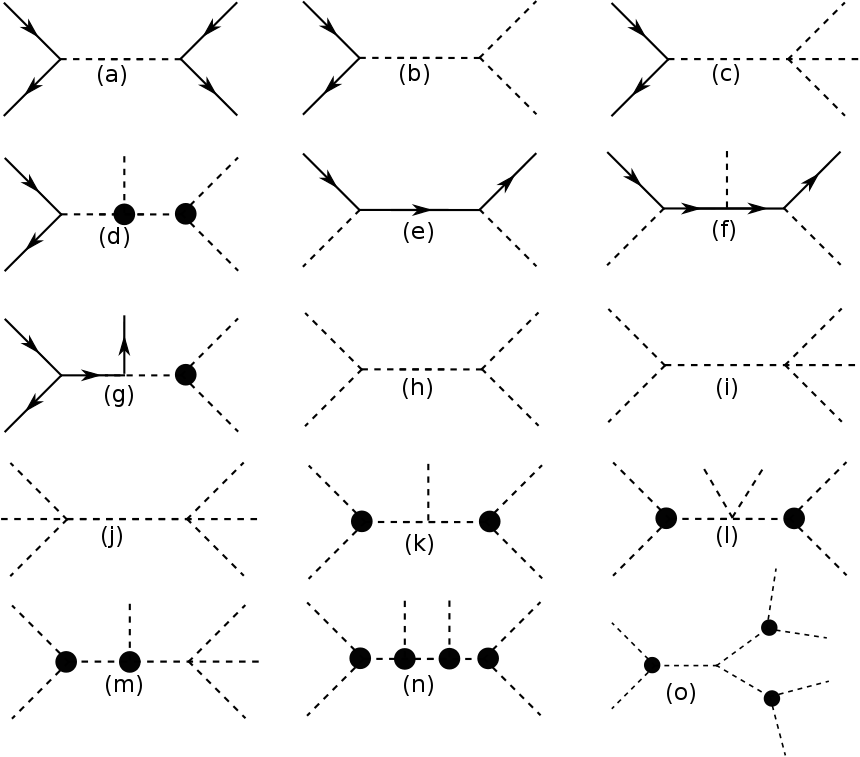}
\caption{\small\label{treegraphs}List of all tree graphs which are suppressed by at most $1/\Lambda^2$ in a general field theory. All~external 
lines corresponds to light fields and internal lines corresponds to heavy fields.}
\end{figure}

 
We assume that the fundamental theory is a gauge theory, that consists of heavy and light fields. All Standard Model and dark matter particles
belong to the second category. Integrating out all heavy fields 
we obtain infinite series of higher dimensional effective operators that are suppressed by increasing powers of 
a high-energy scale~$\Lambda$. If we assume that the SM gauge symmetry $SU(3)_C\times SU(2)_L\times U(1)_Y$ is
a subgroup of the fundamental group, then all these operators are invariant with respect to it.
It can be determined which effective 
operators are generated at tree level, and which at loop level \cite{artz}. The latter are suppressed by factor $\frac{1}{16\pi^2}$, 
therefore they are more difficult to detect.
Possible vertices (fig.~\ref{vertices}) of the underlying theory can be put together in all possible ways to find a complete list of 
tree graphs (fig.~\ref{treegraphs}) that can produce effective operators up to dimension six. Solid lines represent fermions and dashed lines 
represents scalars or vectors. Dotted vertices has coupling constant proportional to heavy scale~$\Lambda$. 
Each internal boson line gives a factor $(p^2-\Lambda^2)^{-1}$ that for momenta $p\ll\Lambda$ reduces to~$\Lambda^{-2}$. 
Fermionic lines give a factor~$\Lambda^{-1}$. 

In order to determine, if an operator can be generated with one of the tree graphs (Fig.~\ref{treegraphs}), it is essential to 
find which vertices between light and heavy fields are allowed in the fundamental theory \cite{artz}. 
$F^a_\mu$ will denote all vectors. $\varphi,\psi,A^a_\mu$ will refer to $SM\nu_R$ fields: scalar, fermions and vectors respectively.
Others field are denoted by $\Phi,\Psi,V^a_\mu$, if they are written with subscript $h$ it means that they are heavy and can be integrated out.
All these fields have canonical form of kinetic terms and refer to mass eigenstates.
\begin{itemize}

 \item Vector interactions

Vector-vector interactions originate from kinetic term $F^a_{\mu\nu}F^{a\mu\nu}$, 
where field tensor read as $F^a_{\mu\nu}=\partial_\mu F^a_\nu - \partial_\nu F^a_\nu - g f_{abc} F^b_\mu F^c_\nu$. 
There are terms $f_{abc} F^a F^b F^c$ with one derivative and terms $f_{abe}f_{cde} F^a F^b F^c F^d$~without. Lie algebra of the fundamental 
gauge group can be divided into generators of the $SM$ symmetry $T_A$ having indices denoted with capital letters and a set of 
other generators $T_a$.  SM gauge group forms
a subgroup of the fundamental group, therefore commutators $[T_A,T_B]$ belongs to this subalgebra. It means, that
structure constants with exactly two indices of SM generators vanish $f_{ABc}=0$. Structure constants for other combination of
indices can be non-zero, therefore vertices $AAV$ and $AAAV$ are absent and 
$AAA, AVV, VVV, AAAA, AAVV, AVVV, VVVV$ are allowed.
\item Scalar interactions

Vertex $\varphi^3$ is not allowed, because it breaks $SU(2)_L$ symmetry. Other three-scalars terms $\Phi\Phi\Phi, \Phi\Phi\varphi, \Phi \varphi^2$ are
possible if there are heavy $SU(2)_L$ singlets and doublets. They can have a coupling proportional to heavy scale $\Lambda$, if there are heavy scalars
with non-zero VEV. All four-scalars vertices are allowed.
\item Vector-fermion interactions
They originate from kinetic term for fermion multiplets $\Pi_i$ in the fundamental theory
\begin{equation}
 \sum_{i}\bar{\Pi}_i \slashed{D} \Pi_i.
\end{equation}
Only $\Psi\psi A$ is forbidden, because it is proportional to $SM$ generator $T_A$. When SM gauge group is unbroken it
cannot mix SM fields with others in the fermion multiplet $\Pi_i$.
\item Scalar-fermion interactions

All Yukawa interactions terms are allowed.
\item Scalar-vector interactions

They come from kinetic term for scalar multiplet $\Omega_i$.
\begin{equation}
 \sum_i (D_\mu \Omega_i)^\dagger (D^\mu \Omega_i)
\end{equation}
When a multiplet $\Omega_i$ consists of non-standard fields $\Phi$ with zero VEV or standard scalar field $\varphi$, then all interactions between
scalars and vectors are allowed. If both types of fields are included in $\Omega_i$ then terms 
$\partial\varphi \Phi A, \varphi\partial\Phi A, \varphi\Phi A A$ are forbidden because they contain a~$SM$ generator $T_A$, which
cannot mix $\varphi$ with non-standard fields $\Phi$. Terms with one scalar field and two vectors appear when $\Omega_i$ has a heavy VEV.
They have structure $\Omega^\dagger_i T T v_h F F$, where $v_h=\langle 0 | \Phi | 0 \rangle$. When the SM symmetry remains unbroken $T_A v_h=0$
and vertices $\varphi A A$ and $\Phi A A$, $\varphi X A$ and $\Phi A X$ are forbidden. $\varphi XX$ is absent.
 
To sum up the forbidden vertices are: 
\begin{equation}
\begin{split}
 \varphi^3,\;\;\;\varphi\Phi AA,\;\;\; \varphi \Phi A,\;\;\;\varphi AX, \;\;\;\varphi AA,\;\;\;\Phi AA, \;\;\; \\ \Phi AX, \;\;\;\varphi XX,\;\;\;
AAX,\;\;\; AAAX,\;\;\;\;\Psi\psi A.
\end{split}
\end{equation}
\end{itemize}

%
%

\subsection{Effective operators suppressed by factor $1/\Lambda$}
Here we collect dim 5 operators and verify if they are tree- or loop-generated.
\vspace{1cm}

\textsc{Tree:}
\begin{itemize}
\item $\Psi^T C \Psi \varphi^\dagger \varphi$

It can be generated by the graph (b) from the list (Fig. \ref{treegraphs}) with vertices $\Psi^T C \Psi \varphi_h$ and  
$\varphi^\dagger \varphi \varphi_h$ from high-energy theory, where $\varphi_h$ is a heavy scalar field.
\item $\nu^T_{Rp} C \nu_{Rq} \Phi^2$

It can be generated at tree level in the same way as operator above with vertices $\nu^T_{Rp} C \nu_{Rq} \varphi_h$ and 
$\Phi^2 \varphi_h$.
\end{itemize}

\textsc{Loop:}
\begin{itemize}
 \item $\overset{(\sim)}{B}_{\mu\nu} \Psi^T C \sigma^{\mu\nu} \Psi$

This operator can be generated through one loop graph with interactions $\overset{(\sim)}B_{\mu\nu}X^{\mu}_{1h} {X}^{\nu}_{2h}$,
$\Psi^T C \gamma_\mu\psi_h X^\mu_{1h}$ and $\bar{\psi}_h\gamma_\nu\Psi X^\nu_{2h}$.

 \begin{fmffile}{BBF}
	        \begin{fmfgraph*}(150,80)
	            \fmfleft{i}
	            \fmfright{j,k}
	            \fmf{photon}{i,v1}
		    \fmf{photon,tension=.7,label=$X^\mu_{1h}$,left=0.5}{v1,v2}
		    \fmf{photon,tension=.7,label=$X^\nu_{2h}$,right=0.5}{v1,v3}
		    \fmf{fermion}{j,v3}
		    \fmf{fermion,tension=.7,label=$\psi_h$}{v3,v2}
		    \fmf{fermion}{v2,k}
	        \end{fmfgraph*}
	    \end{fmffile}

It is worth to notice, that there are no tree graphs with three external lines.

\subsection{Effective operators suppressed by factor $1/\Lambda^2$}
In this section we collect operators, that are suppressed by $1/\Lambda^2$ and check how they could be generated.

\textsc{Tree:}
 \item $\varphi^\dagger\varphi \Phi^4$

It can be easily generated by the graph (j) and $\varphi^\dagger\varphi_h \Phi^2$ interaction.

 \item $(\varphi^\dagger\varphi)^2 \Phi^2$

The same way as above  with $\varphi^\dagger\varphi \varphi_h\Phi$ interaction.

 \item $\varphi^\dagger\varphi \partial_\mu \Phi \partial^\mu \Phi$

It can be generated by (h) and interaction $\partial_\mu\Phi \varphi X^\mu_h$, where $X^\mu_h$ is a heavy gauge vector.

 \item $(D_\mu \varphi)^\dagger D^\mu \varphi \Phi^2$

It can be generated by (h) and interaction $D_\mu\varphi\Phi X_h^{\mu}$.

 \item SM Yukawa interactions $\times \Phi^2$

They can be generated by (f) and interactions $\bar{\psi}_h{\psi}\Phi$, $\bar{\psi}_h\psi_h\varphi$ and $\bar{\psi}\psi_h\varphi$.

 \item $\nu^T_{Rp} C \nu_{Rq} \Psi^T C \Psi$ and $\nu^T_{Rp} C \nu_{Rq} \bar{\Psi}C\bar{\Psi}^T$

They can be generated by (a) and interactions $\nu^T_{Rp} C \nu_{Rq} \varphi_h$ and $\Psi^T C \Psi \varphi_h$ or $\bar{\Psi} C \bar{\Psi}^T \varphi_h$ for
the~latter operator.

\item $\bar{\psi} \gamma_\mu \psi \bar{\Psi} \gamma^\mu \Psi$

It can be generated by (a) and interactions $\bar{\psi} \gamma_\mu \psi  X^\mu_h$ and $ \bar{\Psi} \gamma_\mu \Psi X^\mu_h$.	

\item $\nu^T_{Rp} \sigma_{\mu\nu} \nu_{Rq} \Psi^T C \sigma^{\mu\nu} \Psi$ and $\nu^T_{Rp} \sigma_{\mu\nu} \nu_{Rq} \bar{\Psi} \sigma^{\mu\nu} C\bar{\Psi}^T$ 

Using Fierz identity (\ref{fierztensor}) we obtain:
\begin{equation}
 \nu^T_{Rp} C \sigma_{\mu\nu} \nu_{Rq} \Psi^T C \sigma^{\mu\nu} \Psi = -4 \nu^T_{Rp} C \nu_{Rq} \Psi^T C \Psi - 8\nu^T_{Rp} C \Psi \Psi^T C \nu_{Rq}
\end{equation}
\begin{equation}
 \nu^T_{Rp} C \sigma_{\mu\nu} \nu_{Rq} \bar{\Psi}  \sigma^{\mu\nu} C\bar{\Psi}^T = 
-4 \nu^T_{Rp} C \nu_{Rq} \bar{\Psi} C \bar{\Psi}^T + 8\nu^T_{Rp}\bar{\Psi}^T \bar{\Psi} \nu_{Rq}
\end{equation}

All such operators can be tree generated by (a) and an interaction of fermionic currents with $X_h$ or $\varphi_h$. 

\item $\varphi^\dagger \overleftrightarrow{D}_\mu \varphi \psi \bar{\Psi} \gamma^\mu \Psi$

It can be generated by (b) and interactions $\varphi^\dagger \overleftrightarrow{D}_\mu \varphi X^\mu_h$ and $\bar{\Psi} \gamma^\mu \Psi X^\mu_h$.

\item $\varphi^\dagger\varphi V_\mu V^\mu$

Assuming that $V_\mu$ has mass due to heavy Higgs $h$, this operator can be generated by (h) with vertices
$g^2fhV_\mu V^\mu$ (\ref{higgsint}) and $\varphi^\dagger\varphi h f$, $g\sim v\Lambda^{-1}$ and $f\sim \Lambda$. Therefore
after integrating heavy Higgs it is suppressed by factor $1/\Lambda^2$.

In the Stuckelberg case it is loop-generated by interaction with scalar as in (\ref{stuckscal}). The vertices are
$V_\mu\varphi_h^*\overleftrightarrow{D}_\mu \varphi_h$ and $\varphi_h^*\varphi_h\varphi^\dagger\varphi$.

 \begin{fmffile}{stuck}
	        \begin{fmfgraph*}(150,90)
	            \fmfleft{i,j}
	            \fmfright{k,l}
	            \fmf{photon}{i,v1}
		    \fmf{photon}{j,v2}
		    \fmf{scalar,tension=0.7,label=$\varphi_h$,l.side=left}{v1,v2}
		    \fmf{scalar,tension=0.7,label=$\varphi_h$,l.side=left}{v3,v1}
		    \fmf{scalar,tension=0.7,label=$\varphi_h$,l.side=left}{v2,v3}
		    \fmf{dashes}{v3,k}
		    \fmf{dashes}{v3,l}
	        \end{fmfgraph*}
\end{fmffile}

\end{itemize}

\textsc{Loop:}
\begin{itemize}

\item $\overset{(\sim)}{X}_{\mu\nu} X^{\mu\nu} \Phi^2$ 

It could be tree-generated only by (h), but there are no vertices that join two vectors with derivatives 
or one vector with derivative and scalar field with a heavy field. On the loop level it follows from graph with vertices 
$\overset{(\sim)}{X}_{\mu\nu}X^{\mu}_{1h} {X}^{\nu}_{2h}$, ${X}_{\mu\nu}X^{\mu}_{1h} {X}^{\nu}_{2h}$, $\Phi^2X^{\nu}_{1h} {X}_{1h \nu}$.

 \begin{fmffile}{XXP}
	        \begin{fmfgraph*}(150,70)
	            \fmfleft{i,j}
	            \fmfright{k,l}
	            \fmf{photon}{i,v1}
		    \fmf{photon}{j,v2}
		    \fmf{photon,tension=.7,label=$X^\mu_{1h}$,l.side=left}{v1,v2}
		    \fmf{photon,tension=.7,label=$X^\nu_{2h}$,l.side=right}{v1,v3}
		    \fmf{photon,tension=.7,label=$X_{2h\nu}$,l.side=left}{v2,v3}
		    \fmf{dashes}{v3,k}
		    \fmf{dashes}{v3,l}
	        \end{fmfgraph*}
\end{fmffile}

\item $\varphi^\dagger\varphi \overset{(\sim)}V_{\mu\nu}V^{\mu\nu}$

The generation of this operator is analogous to the previous one. It is generated by loop graph with vertices: 
$\overset{(\sim)}{V}_{\mu\nu}X^{\mu}_{1h} {X}^{\nu}_{2h}$, ${V}_{\mu\nu}X^{\mu}_{1h} {X}^{\nu}_{2h}$, $\varphi^\dagger\varphi X^{\nu}_{1h} {X}_{1h \nu}$.

\end{itemize}

In the end, we have to recall that some operators were eliminated by the equation of motion and therefore are absent on the list above.
However, if any of those operators can be generated by tree-graph, then its unsuppressed coefficient will be added to effective coupling of operators 
to which it was reduced \cite{artz}. In our case operators 
$\overset{(\sim)}{X}_{\mu\nu} X^{\mu\nu} \Phi^2$, $\varphi^\dagger\varphi \overset{(\sim)}V_{\mu\nu}V^{\mu\nu}$ 
do not arise in any operator elimination. The operator $\overset{(\sim)}{B}_{\mu\nu} \Psi^T \sigma^{\mu\nu} \Psi$ (\ref{bPsP}) occurred in 
the elimination of (\ref{longest}), but this operator also needs three external lines in the graph and cannot be generated at tree level. 
The above reasoning also does not change the importance of $\varphi^\dagger\varphi V_\mu V^\mu$ within the Stuckelberg model.

\chapter*{Conclusions}
\addcontentsline{toc}{chapter}{Conclusions}

Empirical facts combined with theoretical predictions clearly indicate, that dark matter has significant share in the energy density of the Universe.
Nevertheless, despite many experimental efforts, the nature of dark matter is still unknown. If we hope to detect dark matter particles, they ought
to interact with the known particles also non-gravitationally. Therefore, while we do not know the correct model of beyond-SM physics, it is useful to 
analyse the problem in the model independent manner and find possible interactions of the Standard Model fields with dark matter candidates.

The thesis presents a classification of interactions between the Standard Model with right-chiral neutrinos ($SM\nu_R$) 
and the dark sector ($DM$) that consists of a real scalar $\Phi$, left and right chiral fermions $\Psi_L$, $\Psi_R$ and a massive vector field $V_\mu$, 
which are singlets of $SU(3)_C\times SU(2)_L\times~U(1)_Y$. In~order to determine the interactions, we obtained full set of gauge singlet operators in 
the~$SM\nu_R$ up to dimension~4 (tab.~\ref{tablesm}). We would like to emphasize that this list can be used to analyse general extensions of the $SM\nu_R$
with fields that are singlets of its gauge group. We found also a similar set of operators for the dark matter fields (tab. \ref{tabledm}),
that is general for scalar and fermions. In the case of a vector field, we~assumed that its
mass should be generated within a renormalizable theory. Therefore we determined only those operators containing $V_\mu$,
that could arise, when the vector mass is generated via the Stuckelberg or the~Higgs mechanism.

The number of operators in the renormalizable models of $SM\nu_R\times DM$ interactions is substantially limited by the maximal
dimension of operators, which is 4 and the stability conditions we imposed (at least one dark matter particle should be stable). 
We enlist models with scalar Higgs portal, Yukawa interactions and
the kinetic mixing term between $U(1)_Y$ and dark matter vector field tensors. It~should be emphasized that the~vector Higgs portal 
($\varphi^\dagger\varphi V_\mu V^\mu$) is~nonrenormalizable and appears only in the effective Lagrangian, where is suppressed by 
square of the high energy scale~$\Lambda$. 

The main result of the thesis is the full set of effective operators up to dimension 6. It is minimal, in the sense, 
that all operators redundant due to the equations of motion, were eliminated from the effective Lagrangian. We note that under our assumptions, 
namely when the~operators are quadratic in the dark matter fields, the effective Lagrangian with terms suppressed by at most $\Lambda^{-2}$
does not include any operators with two different fields from the dark sector. Our list (tab. \ref{tablesmdm}) contains (up to Hermitian
conjugations) one renormalizable 
operator of dimension 4 ($\Phi^2\varphi^\dagger\varphi$), 7~operators suppressed by $\Lambda^{-1}$ and $35$ suppressed by $\Lambda^{-2}$:
$8$ with $\Phi$, $24$ with $\Psi_L$ or $\Psi_R$ and $3$ with $V_\mu$. The~number of effective operators is not large comparing to total
number of 26 operators in~the~$SM\nu_R$~(\ref{SMLag}) and the dark sector~(\ref{DMLag}) Lagrangians. 

We found that operators containing field tensors cannot be tree-generated 
within a general, underlying gauge theory. 
The same is true for the~vector Higgs portal ($\varphi^\dagger\varphi V_\mu V^\mu$), if the vector mass is generated within the Stuckelberg mechanism.
In~consequence, Wilson coefficients $C_k$ of these operators are suppressed by a factor $\frac{1}{16\pi^2}$. 
Therefore interactions, in which they are involved, are more difficult to detect.

%

\appendix

\chapter{The Fierz identities and four-fermion operators\label{four}}

\begin{table}[h]
\small
\begin{center}
\begin{tabular}{|c|c|c|c|c|c|}\hline
$i\setminus j$ & $S$ & $V$ & $T$ & $A$ & $P$\\\hline
$S$ & $-1/4$ & $-1/4$ & $-1/8$ & $-1/4$ & $-1/4$\\\hline
$V$ & $-1$ &  $1/2$ & $0$ & $-1/2$ & $1$\\\hline
$T$ & $-3$ & $0$ & $1/2$ & $0$ & $-3$\\\hline
$A$ & $-1$ &  $-1/2$ & $0$ & $1/2$ & $1$\\\hline
$P$ & $-1/4$ & $1/4$ & $-1/8$ & $1/4$ & $-1/4$\\\hline
\end{tabular}
\caption{\small\label{fierz}Coefficients $C_{ij}$ for Fierz identities.}
\end{center}
\end{table}
The Fierz identities \cite{nagashima},
\cite{itzykson} are the operations that allow to change the order of fermions in the four-fermion interaction 
\begin{equation}
 \Gamma^{i}=(\bar{\psi}_1 O^{i}\psi_2)( \bar{\psi}_3 O_{i}\psi_4)=\sum_j C_{ij}  (\bar{\psi}_1 O^{i}\psi_4)( \bar{\psi}_3 O_{i}\psi_2),
\label{fierzidentity}
\end{equation}
where $O^i$ are various independent $4\times 4$ matrices:
\begin{equation}
 O^S = \mathbf{1},\;\;\;\; O^V=\gamma^\mu,\;\;\;\; O^T=\sigma^{\mu\nu},\;\;\;\; O^A=\gamma^\mu\gamma^5,\;\;\;\; O^P=\gamma^5
\end{equation}
\begin{equation}
 O_S  = \mathbf{1},\;\;\;\; O_V=\gamma_\mu,\;\;\;\; O_T=\sigma_{\mu\nu},\;\;\;\; O_A=-\gamma_\mu\gamma^5,\;\;\;\; O_P=\gamma^5
\end{equation}
The coefficients $C_{ij}$ are presented in tab. \ref{fierz}. For example products of two vector ($i=V$) or tensor ($i=T$)  currents can be written as
\begin{equation}
\begin{split}
(\bar{\psi}_1 \psi_2)( \bar{\psi}_3 \psi_4)=&
-\frac{1}{4}(\bar{\psi}_1 \psi_4)(\bar{\psi}_3 \psi_2)-\frac{1}{4}(\bar{\psi}_1 \gamma^\mu\psi_4)(\bar{\psi}_3\gamma_\mu \psi_2)
-\frac{1}{8}(\bar{\psi}_1 \sigma^{\mu\nu}\psi_4)(\bar{\psi}_3 \sigma_{\mu\nu} \psi_2)\\
&+\frac{1}{4}(\bar{\psi}_1 \gamma^\mu\gamma^5 \psi_4)(\bar{\psi}_3 \gamma_\mu\gamma^5 \psi_2)-
\frac{1}{4}(\bar{\psi}_1 \gamma^5 \psi_4)(\bar{\psi}_3 \gamma^5 \psi_2),\\
  (\bar{\psi}_1 \sigma^{\mu\nu} \psi_2)( \bar{\psi}_3  \sigma_{\mu\nu}\psi_4)=&
-3(\bar{\psi}_1 \psi_4)(\bar{\psi}_3 \psi_2)
+\frac{1}{2	}(\bar{\psi}_1 \sigma^{\mu\nu}\psi_4)(\bar{\psi}_3 \sigma_{\mu\nu} \psi_2)\\
&-3(\bar{\psi}_1 \gamma^5 \psi_4)(\bar{\psi}_3 \gamma^5 \psi_2).\\
\end{split}
\end{equation}
These identities can be used to express product of tensor currents in terms of scalar currents. In case of Weyl spinors it reads as follows
\begin{equation}
\begin{split}
(\bar{\chi}_1 \sigma^{\mu\nu} \chi_2)( \bar{\chi}_3  \sigma_{\mu\nu}\chi_4)=&-6(\bar{\chi}_1\chi_4)( \bar{\chi}_3 \chi_2)
+\frac{1}{2}(\bar{\chi}_1 \sigma^{\mu\nu} \chi_4)( \bar{\chi}_3  \sigma_{\mu\nu}\chi_2)=\\
&-8(\bar{\chi}_1\chi_4)( \bar{\chi}_3 \chi_2)-4(\bar{\chi}_1\chi_2)( \bar{\chi}_3 \chi_4).
\label{fierztensor}
\end{split}
\end{equation}

Lorentz singlet four-fermion operators have the form $\Gamma^i$ as in (\ref{fierzidentity}). When one is looking for all 
such operators with two fermions from one particle sector and the other two from the second one,
it is enough to take into account the operators with fields from the same sector belonging to one fermionic current. If $\Psi_1$ and $\Psi_4$
in $\Gamma^i$ are fermions from the same sector, then we can put them in one current simply by using Fierz identities. The other possibility,
with $\Psi_1$ and $\Psi_3$ from the same sector, can be reduced to the previous case by the following expression 
\begin{equation}
\bar{\psi}_3 O_i \psi_4 = \psi^\dagger_3 \gamma^0 O_i \psi_4 = -\psi^T_4 O^T_i \gamma^0 \psi^*_3 = -\psi^T_4 CC O^T_i CC \gamma^0 \psi^*_3
=-\overline{\psi_4^c} C O^T_i C \psi^c_3,
\end{equation}
where $C O^T_i C=\pm O_i$, e.g. $C\mathbf{1}C=-\mathbf{1}$, $C\gamma^T_\mu C= \gamma_\mu$,  $C\sigma^T_{\mu\nu} C= \sigma_{\mu\nu}$. For example
\begin{equation}
\begin{split}
& \bar{\nu}_R\Psi_L\bar{\nu}_R\Psi_L=\bar{\nu}_R\Psi_L\overline{\Psi^c_L} \nu^c_R =
-\frac{1}{2}\bar{\nu}_R\nu^c_R\overline{\Psi^c_L}\Psi_L
-\frac{1}{8}\bar{\nu}_R\sigma_{\mu\nu}\nu^c_R\overline{\Psi^c_L}\sigma^{\mu\nu}\Psi_L = \\
&-\frac{1}{2}(\nu^T_R C \nu_R \bar{\Psi}_L C \bar{\Psi}^T_L)^\dagger -\frac{1}{8}(\nu^T_R C \sigma_{\mu\nu}\nu_R 
\bar{\Psi}_L \sigma_{\mu\nu} C \bar{\Psi}^T_L)^\dagger.
\end{split}	
\end{equation}
These are Hermitian conjugations of operators appearing in tab. \ref{tablesmdm}.

\chapter{Useful formulas}

\begin{flushleft}
\begin{equation}
\sigma^{\mu\nu}\partial_\mu =
 \frac{i}{2}(\gamma^{\mu}\gamma^{\nu}-\gamma^{\nu}\gamma^{\mu})\partial_\mu=i(g_{\mu\nu}-\gamma^\nu\gamma^\mu)\partial_\mu=
i(\partial^\nu-\gamma^\nu\slashed{\partial})
 \label{dsigma}
\end{equation}	
\end{flushleft}

\begin{equation}
\begin{split}
\tilde{X}_{\mu\rho}X^{\rho\nu}=&
\frac{1}{2}\varepsilon_{\mu\rho\sigma\delta}X^{\sigma\delta} X^{\rho\nu}=\\
=&\frac{1}{2}{\delta_{\mu}}^{\bar{\nu}}\varepsilon_{\bar{\nu}\rho\sigma\delta}X^{\sigma\delta} X^{\rho\nu}+
\frac{1}{2}\varepsilon_{\mu\bar{\nu}\sigma\delta}X^{\sigma\delta} X^{\bar{\nu}\nu}+
\frac{1}{2}\varepsilon_{\mu\rho\bar{\nu}\delta}X^{\bar{\nu}\delta} X^{\rho\nu}+
\frac{1}{2}\varepsilon_{\mu\rho\delta\bar{\nu}}X^{\delta\bar{\nu}} X^{\rho\nu}\\
=&\frac{1}{8}{\delta_{\mu}}^{\nu}\varepsilon_{\gamma\rho\sigma\delta}X^{\sigma\delta} X^{\rho\gamma}=
-\frac{1}{4}\delta^\nu_\mu\tilde{X}_{\rho\gamma}X^{\rho\gamma},
\end{split}
\label{symten}
\end{equation}
$\bar{\nu}$ means that $\nu$ is excluded from summing. To understand the second equality note that for every non-vanishing term
in the summation all indices in totally antisymmetric tensor must be different, therefore one of them must be equal to $\nu$. 
Last three terms in the second line vanish due to $X^{\nu\nu}=0$ or they contain contraction of antisymmetric and symmetric tensor 
in indices $\rho$ and $\sigma$.

\begin{equation}
 \gamma_\rho\gamma_\mu\gamma_\nu=g_{\rho\mu}\gamma_\nu + g_{\mu\nu}\gamma_\rho- g_{\rho\nu}\gamma_\mu 
+ \gamma_{[\rho}\gamma_\mu\gamma_{\nu]} = \gamma_\rho\gamma_\mu\gamma_\nu=g_{\rho\mu}\gamma_\nu + g_{\mu\nu}\gamma_\rho- g_{\rho\nu}\gamma_\mu 
- i\varepsilon_{\rho\mu\nu\sigma}\gamma^\sigma\gamma_5
\label{ggg}
\end{equation}

\addcontentsline{toc}{chapter}{Bibliography}

\bibliography{PracaMGRvA1}
\bibliographystyle{hunsrt}

\end{document}